\documentclass[aps,prd,amssymb,eqsecnum,showpacs] {revtex4}
\usepackage{graphicx}
\usepackage{psfrag}
\usepackage[mathscr]{eucal}
\usepackage{mathrsfs}

\newcommand{\scri}{\mathscr{I}}

\begin{document}

\title{A Characteristic Extraction Tool for Gravitational Waveforms}

\author{M.~C. Babiuc${}^{1,3}$, B. Szil\'{a}gyi${}^{2,3}$, J. Winicour${}^{3,4}$
and Y. Zlochower${}^{5}$
       }
\affiliation{
${}^{1}$ Department of Physics \\
 Marshall University, Huntington, WV 25755, USA \\
${}^{2}$ Theoretical Astrophysics, California Institute of Technology \\
               Pasadena, CA 91125, USA \\
${}^{3}$ Department of Physics and Astronomy \\
        University of Pittsburgh, Pittsburgh, PA 15260, USA\\
${}^{4}$ Max-Planck-Institut f\" urÄ
         Gravitationsphysik, Albert-Einstein-Institut, 
	 14476 Golm, Germany \\
${}^{5}$ Center for Computational Relativity and Gravitation and 
School of Mathematical Sciences\\
Rochester Institute of Technology, Rochester, NY 14623
}

\begin{abstract}

We develop and calibrate a characteristic waveform extraction tool whose
major improvements and corrections of prior versions allow satisfaction
of the accuracy standards required for advanced LIGO data analysis.
The extraction tool uses a characteristic evolution code to propagate
numerical data on an inner worldtube supplied by a $3+1$ Cauchy evolution
to obtain the gravitational waveform at null infinity. With the new extraction tool,
high accuracy and convergence of the numerical error can be demonstrated for an
inspiral and merger of mass $M$ binary black holes even for an extraction worldtube
radius as small as $R=20M$. The tool provides a means for unambiguous
comparison between waveforms generated by evolution codes based upon
different formulations of the Einstein equations and based upon different numerical
approximations.

\end{abstract}


\maketitle

\section{Introduction}

The strong emission of gravitational waves from the inspiral and merger of
binary black holes has been a dominant motivation for the construction of
gravitational wave observatories. The computation of the precise details of the
waveform by means of numerical simulation is a key theoretical tool to enhance
detection and allow useful scientific interpretation of the gravitational
signal. See~\cite{lindbo} for a review of the accuracy required of numerically
generated waveforms to fully complement the sensitivity of the LIGO~\cite{ligo}
and Virgo~\cite{virgo} observatories.  However, the waveforms are not easy to
extract accurately. The radiation falls off as $1/r$ so that although it
asymptotically dominates near field gravitational effects it is nevertheless
small and can be contaminated by numerical error. It is common practice for the
Cauchy codes used in simulating the binary black hole problem to introduce a
large but finite artificial outer boundary.  A combination of linearized and far
field approximations are then used to extract the waveform from data on a
smaller worldtube, which ideally is causally isolated from the outer boundary. 
Such perturbative wave extraction at a finite distance, rather than at null
infinity which more faithfully represents the idealization of a distant antenna,
introduces systematic errors associated with the effects of gauge,
nonlinearities, nonradiative near fields and back reflection. (See
~\cite{lehnmor,strat} for analyses of waveform errors arising from perturbative
extraction at a finite distance.) An alternative approach called {\it
Cauchy-characteristic extraction} (CCE)~\cite{ccm,cce} provides a fully
nonlinear interface between Cauchy and characteristic codes which utilizes the
characteristic evolution to extend the simulation to null infinity, where the
waveform is computed. An earlier implementation of CCE has recently been applied
to extract waveforms from binary black hole simulations~\cite{reis1,reis2}, from
rotating stellar core collapse~\cite{reis3} and to explore the memory
effect~\cite{reis4}. In this work, we present details and tests of a redesigned
CCE module whose accuracy and efficiency has undergone major improvement. The
module has been designed to provide a standardized waveform extraction tool for
the numerical relativity community which will allow CCE to be readily applied to
a generic Cauchy code. 

The first attempts to simulate collisions of black holes by  Hahn and
Lindquist~\cite{hahn}, and then by Smarr et al~\cite{smarr},  were hampered by
both a lack of computing power and a proper understanding  of the mathematical
formulation of Einstein's equations required for a stable numerical solution.
Their work formed the impetus for the Binary Black Hole Grand Challenge, which
was formed to take advantage of the increasingly powerful computers introduced
in the 1980's. The main results of the Grand Challenge were limited to the
axisymmetric head-on collision of black holes and the gravitational collapse of
rotating matter~\cite{gc}. However, the standard
Arnowitt-Deser-Misner~\cite{adm} formulation of the Einstein equations adopted
by the Grand Challenge had instabilities at the analytic level which limited
more general binary black hole simulations to the premerger stage. Only with new
formulations was a full inspiral and merger successful, first by
Pretorius~\cite{fP05} using the harmonic formulation, and soon after by 
Campanelli et al~\cite{mCcLpMyZ06} and Baker et al~\cite{jBjCdCmKjvM06b}  using
the Baumgarte-Shapiro-Shibabta-Nakamura formulation~\cite{bssn1,bssn2}. Numerous
groups  now have codes which can simulate this binary inspiral problem by
evolving the Cauchy problem for Einstein's equations. 

In CCE,  the Cauchy evolution is used to supply boundary data on a timelike
inner worldtube to carry out a characteristic evolution extending to future null
infinity $\scri^+$, where the waveform can be unambiguously computed using the
geometric methods developed by Bondi et al~\cite{bondi}, Sachs~\cite{sachsr} and
Penrose~\cite{Penrose}. This initial-boundary value problem based upon a
timelike worldtube~\cite{tam} has been implemented as a characteristic evolution
code, the PITT null code~\cite{isaac,highp}, which incorporates a Penrose
compactification of the space-time.  It computes the Bondi news function at
$\scri^+$, which is an invariantly defined complex radiation amplitude $N=
N_{\oplus}+i N_{\otimes}$, whose real and imaginary parts correspond to the time
derivatives $\partial_t h_{\oplus}$ and $\partial_t h_{\otimes}$ of the ``plus''
and ``cross'' polarization modes of the strain incident on a gravitational wave
antenna. The error in the PITT code was tested to be second order convergent in
analytic  testbeds ranging from the perturbative regime~\cite{babiuc05} to
highly nonlinear single black hole spacetimes~\cite{highp}. One of the successes
of the Grand Challenge was  the successful application of the code to generic
single black hole dynamical spacetimes~\cite{stablett,wobb,fiss,mod}. For a
review, see~\cite{livccm}. 

The propagation of gravitational waves to $\scri^+$ from an astrophysically
realistic source using the PITT code has in the past been limited to the
simulation of an imploding neutron star using a fluid dynamic code incorporated
into the characteristic code~\cite{papadop2003,linke}. These simulations were
restricted to the axisymmetric case because of computational demands arising at
the center of the star.  For such systems, CCE offers a way to combine the
strengths of the Cauchy and characteristic approaches. Recently this combined
approach has been applied to extract the waveform from the fully 3-dimensional
collapse of a rotating star~\cite{reis3}.  A global characteristic  simulation
of the full inspiral and merger of a relativistic binary system is not possible
because of the interior caustics formed by gravitational lensing. But the
application of CCE to this system has been shown to be
possible~\cite{reis1,reis2}.

The error in CCE arises from three independent sources: {\bf (I)} the Cauchy
evolution; {\bf (II)} the worldtube module; and {\bf (III)} the characteristic
evolution to $\scri^+$ and the computation of the waveform.

\medskip

{\bf (I)} Errors in the Cauchy evolution can arise from numerical
approximations, improper boundary treatments, extraneous radiation content in
the initial data, instabilities and bugs. Errors introduced at the outer grid
boundary present a special problem for BSSN formulations for which there is no
theoretical understanding of the proper boundary condition. The standard
practice is to extract the waveform at a finite worldtube which is large enough
to justify a far field approximation but which is causally isolated from the
outer boundary during the simulation. For example, perturbative extraction at
$r=100M$  would require that the outer boundary be at $r>500M$ for a $t\approx
400M$ simulation. We have designed the new CCE module so that it  can be applied
to a generic Cauchy code with extraction radius as small as $r=20M$. However,
since any universally applicable extraction module must be designed to be
independent of error introduced by the Cauchy code, the extracted waveform
cannot be any more reliable than the Cauchy code. 

\medskip

{\bf (II)} The main improvement described and tested in this paper is a complete
overhaul of the worldtube module, which converts the output of the Cauchy
evolution to boundary data on an inner worldtube for the characteristic
evolution. The prior version of this module, which was used in the first
applications of CCE to obtain binary black hole
waveforms~\cite{reis1,reis2,reis3}, contained inconsistencies and bugs which
prevented clean convergence tests. We have corrected this worldtube module so
that the present version exhibits clean convergence to which Richardson
extrapolation can be applied to produce waveforms whose numerical error due to
CCE is extremely small. In addition to improvement in consistency and accuracy,
we have also redesigned the module to be more efficient and user friendly.
These revisions are described in Appendix~\ref{sec:rev}.

{\bf (IIII)} In addition to thoroughly scrutinizing the PITT null code for bugs,
we made several major  modifications. In
previous applications requiring very high resolution, such
as the inspiral of a particle into a black hole~\cite{partbh}, there was
excessive short wavelength noise which affected the
quality of the simulation. In addition, in~\cite{reis1,reis2} it was reported
that one of the equations governing calculation of the waveform
at $\scri^+$ had to be linearized
in order to obtain reasonable behavior. These problems have been eliminated
as a result of the modifications described in Appendix~\ref{sec:rev}.

In Sec's.~\ref{sec:chform} and \ref{sec:tests},
we review the formalism underlying characteristic evolution and the computational
structure of the PITT code. We include enough details to make clear
the difficulties underlying extraction of
an accurate waveform at $\scri^+$ and to explain the code modifications that have
been made. We also demonstrate how the use of 4th
order accurate angular derivatives improves the previous test results of CCE presented
in~\cite{strat}. In Sec.~\ref{sec:interface}, we describe the design of the new
worldtube module, how it treats the Cauchy-characteristic interface and how
it can be readily applied to a Cauchy evolution. 

In Sec.~\ref{sec:bbhwaveforms}, we test the new extraction tool on the
Cauchy evolution of the inspiral and merger of two equal-mass, non-spinning black
holes. We show that CCE can now be carried out for
a worldtube radius as small as $20 M$ for a mass $M$ binary system, for which
perturbative extraction would not be meaningful, and which was not
possible with the prior implementation of CCE. Convergence tests now
demonstrate clean second order global accuracy of the evolution variables.

The waveforms are only first order accurate as a result of
the asymptotic limits required at $\scri^+$.  However,
the clean first order convergence of the waveform now allows application of
Richardson extrapolation to obtain higher order accuracy. In this way,
in Sec.~\ref{sec:rich}, we construct a third order accurate waveform,
which was not possible with earlier versions of CCE.

The ability to apply Richardson extrapolation
to CCE waveforms makes it possible to show that their 
numerical error satisfies the standards required for application to advanced
LIGO data analysis. The first  derivation~\cite{flanhugh} of
the accuracy required for numerically generated black hole waveforms 
to be useful as templates for LIGO was carried out in
the frequency domain, in which the interferometer noise spectrum is calibrated.
There are two separate criteria: one ensures that
the error in the model waveform does not impact wave detection and the other
ensures that the error does not impact the scientific content of the signal.
These criteria both depend upon the noise spectrum of the detector
in a way not easily applied to a numerical simulation. This has recently prompted
a translation of these requirements into the time domain in which
the waveforms are computed~\cite{lindbo,lindob,lind}, so that they can
be readily enforced in practice. In Sec.~\ref{sec:crit} we show that the numerical
error introduced by CCE  satisfies these time domain criteria for an
advanced LIGO detector. We also analyze the error introduced by the choice of
initial data, which has a dependency upon the size of the extraction worldtube.

\section{Characteristic Formalism}
\label{sec:chform}

The characteristic formalism is based upon a family of outgoing null
hypersurfaces emanating from an inner worldtube and extending to
infinity where they foliate $\scri^+$ into spherical slices. 
We let $u$ label these hypersurfaces, $x^A$ $(A=2,3)$ be angular coordinates
which label the null rays and $r$ be a surface area coordinate. In the
resulting $x^\alpha=(u,r,x^A)$ coordinates, the metric takes the Bondi-Sachs
form~\cite{bondi,sachsr}
\begin{eqnarray}
   ds^2 & = & -\left(e^{2\beta}\frac{V}{r} -r^2h_{AB}U^AU^B\right)du^2
        -2e^{2\beta}dudr -2r^2 h_{AB}U^Bdudx^A \nonumber \\
        & + & r^2h_{AB}dx^Adx^B,    
	\label{eq:bmet}
\end{eqnarray}
where $h^{AB}h_{BC}=\delta^A_C$ and
$det(h_{AB})=det(q_{AB})$, with $q_{AB}$ a unit sphere metric.  In
analyzing the Einstein equations, we also use the intermediate variable
\begin{equation}
     Q_A = r^2 e^{-2\,\beta} h_{AB} U^B_{,r}.
\end{equation}
Because the Bondi variable $V=O(r^2)$ at $\scri^+$, the code is written in
terms or the renormalized variable $W=(V-r)/r^2$. Here $W=0$ for the Minkowski
metric in null spherical coordinates. 

The PITT null code employs a spherical grid based upon an auxiliary unit sphere
metric $q_{AB}$, with associated complex dyad $q_A$ satisfying $ q_{AB}
=\frac{1}{2}\left(q_A \bar q_B+\bar q_Aq_B\right)$. The Bondi-Sachs metric
$h_{AB}$ induced on the spherical cross-sections can then be represented by its
dyad component $J=h_{AB}q^Aq^B /2$, with the spherically symmetric case
characterized by $J=0$. The fully nonlinear $h_{AB}$ is uniquely determined by
$J$, which is the principle evolution variable. The determinant condition
implies that the dyad component $K=h_{AB}q^A \bar q^B /2$ is determined by
$1=K^2-J\bar J$. We also introduce spin-weighted fields $U=U^Aq_A$ and
$Q=Q_Aq^A$, as well as the complex spin-weight operators $\eth$ and $\bar
\eth$~\cite{newp} which represent the angular derivatives. Refer to~\cite{eth}
for details regarding numerical implementation.

In this formalism, the Einstein equations decompose into hypersurface
equations, evolution equations and conservation conditions on the inner
worldtube. As described in more detail in~\cite{tam,cce}, the
hypersurface equations take the form
\begin{eqnarray}
      \beta_{,r} &=& N_\beta[J], 
   \label{eq:beta} \\
     (r^2 Q)_{,r}  &=& -r^2 (\bar \eth J + \eth K)_{,r}
                +2r^4\eth \left(r^{-2}\beta\right)_{,r} + N_Q[J,\beta], 
     \label{eq:wq} \\
          U_{,r}  &=& r^{-2}e^{2\beta}Q +N_U[J,\beta,Q], 
     \label{eq:wua} \\
      V_{,r} &=& \frac{1}{2} e^{2\beta}{\cal R} 
- e^{\beta} \eth \bar \eth e^{\beta}
+ \frac{1}{4} r^{-2} \left(r^4
                           \left(\eth \bar U +\bar \eth U \right)
                     \right)_{,r} + N_W[J,\beta,Q,U],
 \label{eq:ww}
\end{eqnarray}
where
\begin{equation}
{\cal R} =2 K - \eth \bar \eth K + \frac{1}{2}(\bar \eth^2 J + \eth^2 \bar J)
          +\frac{1}{4K}(\bar \eth \bar J \eth J - \bar \eth J \eth \bar J)
     \label{eq:calR}
\end{equation}
is the curvature scalar of the 2-metric $h_{AB}$. 
The evolution equation for $J$ takes the form
\begin{eqnarray}
    && 2 \left(rJ\right)_{,ur}
    - \left(r^{-1}V\left(rJ\right)_{,r}\right)_{,r} = \nonumber \\
    && -r^{-1} \left(r^2\eth U\right)_{,r}
    + 2 r^{-1} e^{\beta} \eth^2 e^{\beta}- \left(r^{-1} V \right)_{,r} J
    + N_J[J,J_{,u},\beta,Q,U,W],
    \label{eq:wev}
\end{eqnarray}
where $N_\beta[J]$, $N_Q[J,\beta]$, $N_U[J,\beta,Q]$, $N_W[J,\beta,Q,U]$ and
$N_J[J,J_{,u},\beta,Q,U,W]$ are nonlinear terms which vanish for spherical symmetry
and can be constructed from the hypersurface values of the variables appearing in
their argument. Expressions for these nonlinear terms as complex spin-weighted
fields and a discussion of the conservation conditions are given in~\cite{cce}.
The hypersurface equations have a hierarchical structure in the order
$[J,\beta,Q,U,W]$ such that the right hand sides, e..g. $N_\beta[J]$ only depend
upon previous variables and their derivatives intrinsic to the hypersurface.

The finite difference grid used in the code is based upon the compactified
radial coordinate
\begin{equation}
      x=\frac{r}{R_E +r}
\label{eq:compx}
\end{equation}
so that $x=1$ at $\scri^+$. Here $R_E$ is a parameter which in the CCE module is
chosen as the radius of the extraction worldtube as determined by
$R^2=\delta_{ij}x^i x^j$ in terms of the Cartesian coordinates $x^i$ used in the
Cauchy evolution code. 

The auxiliary variables
\begin{equation}
     \nu =\bar \eth J \, , \quad {\cal B}=\eth \beta \, , \quad k=\eth K
\label{eq:aux}
\end{equation}
are also introduced to eliminate all second angular derivatives.
In certain applications this has been found to give rise to increased
accuracy by suppressing short wavelength error~\cite{gomezfo}. 

The finite difference scheme for integrating the hypersurface and evolution
equations has been described in~\cite{highp,gomezfo,luisdis}. Except for the
start-up procedure described in Sec.~\ref{sec:interface}, we follow this scheme
with two modifications. First, the finite difference approximation for the
$\eth$-operators is increased from 2nd order to 4th order accuracy. This can be
expected to give better angular resolution but does not affect the overall 2nd
order accuracy implied by the radial and time integration schemes. Second, when
rewritten in terms of the compactified $x$-coordinate, the hypersurface
equations for $Q$ and $W$ take the form
\begin{equation}
     x(1-x)\partial_x F + 2F = RHS
\label{eq:sing}
\end{equation}
where the right hand side is regular at $\scri^+$. In order to deal
with the degeneracy of this equation at $x=1$, we rewrite (\ref{eq:sing}) in the form
\begin{equation}
     \frac {\partial (r^2 F)} {\partial (r^2)}= \frac {RHS}{2}
\label{eq:singfd}
\end{equation}
and construct a centered finite difference approximation with respect to
$r^2$. Expressed in terms of the grid $x_i = x_{i-i} +\Delta x$, this leads to 
\begin{equation}
      F_i =\bigg ( \frac {x_{i-1}(1-x_i)}{x_i(1-x_{i-1})} \bigg )^2 F_{i-1} 
            +\frac{\Delta x(x_i+x_{i-1}-2x_i x_{i-1})}{x_i^2(1-x_{i-1})^2}
          \frac{RHS}{2} 
\label{eq:asymqw}    
\end{equation}
which enforces the correct asymptotic limit $F|_{x=1} = RHS/2$ when $RHS$ is
constant near $\scri^+$. In practice, the variation of $RHS$ implies that this
limit is only enforced to first order accuracy when $RHS$ is evaluated by the
mid-point rule. This is consistent with global second order accuracy of $Q$ and
$W$ when the numerical error is measured by an $L_2$-norm over the hypersurface,
but only first order accuracy can be expected for their values at $\scri^+$.
However, the asymptotic values of $Q$ and $W$ do not enter directly into the
calculation of the waveform at $\scri^+$.

\subsection{Waveforms at $\scri^+$}
\label{sec:waveforms} 

For technical simplicity, the theoretical derivation of the waveform at infinity
is best presented in terms of an inverse surface-area coordinate $\ell=1/r$,
where  $\ell=0$ at $\scri^+$. In the resulting $x^\mu=(u,\ell,x^A)$ conformal
Bondi coordinates, the physical space-time metric $g_{\mu\nu}$ has the conformal
compactification $\hat g_{\mu\nu}=\ell^{2} g_{\mu\nu}$, where $\hat g_{\mu\nu}$
is smooth at $\scri^+$ and, referring to (\ref{eq:bmet}), takes the
form~\cite{tam}
\begin{equation}
   \hat g_{\mu\nu}dx^\mu dx^\nu= 
           -\left(e^{2\beta}V \ell^3 -h_{AB}U^AU^B\right)du^2
        +2e^{2\beta}dud\ell -2 h_{AB}U^Bdudx^A + h_{AB}dx^Adx^B.
   \label{eq:lmet}
\end{equation}

As described in~\cite{strat}, the Bondi news function $N(u,x^A)$ and the
Newman-Penrose Weyl tensor component~\cite{NP}
$$\Psi_4^0(u,x^A)=\lim_{r\rightarrow \infty} r \psi_4
$$ 
which describe the waveform are both determined by the
asymptotic limit at $\scri^+$ of the tensor  field
\begin{equation}
 \hat \Sigma_{\mu\nu} = \frac{1}{\ell}(\hat \nabla_\mu\hat \nabla_\nu   
  -\frac{1}{4}\hat g_{\mu\nu} \hat \nabla^\alpha\hat \nabla_\alpha)\ell.
\label{eq:Sigma}
\end{equation}
This limit is  constructed from the leading coefficients in an
expansion of the metric in powers of $\ell$. We thus write
\begin{equation}
   h_{AB}= H_{AB}+\ell c_{AB}+O(\ell^2).
\end{equation}
Conditions on the asymptotic expansion of the remaining components of
the metric follow from the Einstein equations:
\begin{equation}
    \beta=H+ O(\ell^2) ,
\end{equation}
\begin{equation}
    U^A= L^A+2\ell e^{2H} H^{AB}D_B H+O(\ell^2) 
\end{equation}
and 
\begin{equation}
    \ell^2 V= D_A L^A
     +\ell (e^{2H}{\cal R}/2 +D_A D^A e^{2H})+O(\ell^2),
\end{equation}
where $H$ and $L$ are the asymptotic limits of $\beta$ and $U$ and
where ${\cal R}$ and $D_A$ are the 2-dimensional curvature scalar and
covariant derivative associated with $H_{AB}$. 

The expansion coefficients $H$, $H_{AB}$, $c_{AB}$ and $L^A$ (all functions of
$u$ and $x^A$) completely determine the radiation field. One can further
specialize the Bondi coordinates to be {\em inertial} at $\scri^+$, i.e. have
Minkowski form, in which case $H=L^A=0$, $H_{AB}=q_{AB}$ (the unit sphere
metric) so that the radiation field is completely determined by $c_{AB}$.
However, the characteristic extraction of the waveform is carried out in
computational coordinates determined by the Cauchy data on the extraction
worldtube so that this inertial simplification cannot be assumed. 

In order to compute the Bondi news function in the $\hat g_{\mu\nu}$
frame, it is necessary to determine the conformal factor $\omega$
relating  $H_{AB}$ to a unit sphere metric $Q_{AB}$, i.e.  to an
inertial conformal Bondi frame~\cite{tam} satisfying
\begin{equation}
          Q_{AB}=\omega^2H_{AB}.
\label{eq:unsph}
\end{equation}
(See~\cite{quad} for a discussion of how the news in an arbitrary
conformal frame is related to its expression in this inertial Bondi
frame.) We can determine $\omega$ by solving the elliptic equation
governing the conformal transformation of the curvature scalar
(\ref{eq:calR}) to a unit sphere geometry,
\begin{equation}
     {\cal R}=2(\omega^2+H^{AB}D_A D_B \log \omega).
\label{eq:conf}
\end{equation}
Equation (\ref{eq:conf}) need only be solved at the initial time. Then
the  geometrical properties of $\scri^+$ determines the time
dependence of $\omega$ according to
\begin{equation}
     2\hat n^\alpha \partial_{\alpha} \log \omega
    =-e^{-2H}D_AL^A,
\label{eq:omegadot}
\end{equation}
where $\hat n^\alpha =\hat g^{\alpha\beta}\nabla_\beta \ell$ is the null vector
tangent to the generators of $\scri^+$. We use (\ref{eq:omegadot}) to evolve
$\omega$ along the generators of $\scri^+$ given a solution of (\ref{eq:conf})
as initial condition.

The news function $N(u,x^A)$ is first computed by the code in terms of the
computational coordinates $(u,x^A)$, as opposed to the inertial coordinates
$(\tilde u,y^A)$ on $\scri^+$ corresponding to an idealized distant observatory.
The transformation to inertial coordinates proceeds by introducing the
conformally rescaled metric $\tilde g_{\mu\nu} = \omega^2 \hat g_{\mu\nu}$ in
which the cross-sections of $\scri^+$ have unit sphere geometry, in accord with
(\ref{eq:unsph}). The rescaled null vector $\tilde n^\nu = \omega^{-1} \hat
n^\nu$ is then the generator of the inertial time translation on $\scri^+$, i.e.
$\tilde n^\nu \partial_\nu = \partial_{\tilde u}$. The inertial coordinates thus
satisfy the propagation equations
\begin{equation} 
      \hat n^\nu \partial_\nu \tilde u = \omega \, , \quad
        \hat n^\nu \partial_\nu y^A =0,
\label{eq:inertialc}
\end{equation} 
where $\hat n^\nu\partial_\nu =e^{-2H}(\partial_u + L^A
\partial_{x^A})$ in terms of the computational coordinates. The
inertial coordinates are obtained by integrating (\ref{eq:inertialc}),
thus establishing a second pair of stereographic grid patches
corresponding to $y^A$. Then the news function is transformed into
$N(\tilde u, y^A)$. (More precisely, we should write $\tilde N(\tilde
u, y^A) = \hat N(u,x^A)$ to distinguish the functional form of the news
in the different coordinates but we forgo this complication of notation.)

In addition, in order for the real and imaginary parts of $N$ to
correspond to the ``plus'' and ``cross'' polarization modes of a
distant observatory, we need the proper choice of complex polarization
vector ${\cal Q}^{\beta}$, which in the inertial coordinates is related
to the unit sphere metric on $\scri^+$ by $Q^{AB} =({\cal Q}^A \bar
{\cal Q}^B +\bar {\cal Q}^A  {\cal Q}^B)/2$. We fix the spin rotation
freedom ${\cal Q}^{\beta} \rightarrow e^{-i\eta}{\cal Q}^{\beta}$ by
requiring ${\tilde n}^{\nu}{\tilde \nabla}_{\nu} {\cal
Q}^{\beta}=O(\Omega)$, so that the polarization frame is parallel
propagated along the inertial time flow on $\scri^+$. This fixes the
polarization modes determined by the real and imaginary parts of the
news to correspond to those of inertial observers at $\scri^+$. In
order to carry this out in the computational frame we introduce the
dyad decomposition  $H^{AB}=(F^A{\bar F}^B+{\bar F}^A  F^B)/2$ where
\begin{equation}
   F^A  = q^A  \sqrt{ \frac{(K+1)}{2 }  }
          -\bar q^A  J \sqrt{ 1 \over 2(K+1)} .
\end{equation}
We then set ${\cal Q}^{\beta}=e^{-i\delta}\omega^{-1}F^\beta 
+\lambda {\tilde n}^{\beta}$, where $F^\alpha:=(0,0,F^A)$. The
requirement of an inertial polarization frame,
${\tilde n}^{\nu}{\tilde \nabla}_{\nu} {\cal Q}^{\beta}
=O(\Omega)$, then determines the time dependence of the phase
$\delta$ according to
\begin{equation}
    2i(\partial_u +L^A\partial_A)\delta = D_A L^A
     +H_{AC} \bar F^C ( (\partial_u +L^B \partial_B) F^A
             - F^B \partial_B L^A) .
    \label{eq:evphase}
\end{equation}
The Bondi news  now takes the explicit form
\begin{equation}
    N={1\over 4}e^{-2i \delta}\omega^{-2}e^{-2H}F^A F^B
       \{(\partial_u+{\pounds_L})c_{AB}-{1\over 2}c_{AB} D_C L^C
        +2\omega D_A[\omega^{-2}D_B(\omega e^{2H})]\},
     \label{eq:news}
\end{equation}
where $\pounds_L$ denotes the Lie derivative with respect to $L^A$.

In the inertial Bondi coordinates, the expression for the news function
reduces to the simple form
\begin{equation}
    N={1\over 4}{\cal Q}^A {\cal Q}^B \partial_u c_{AB}.
     \label{eq:inews}
\end{equation}
However, the general form (\ref{eq:news}) must be used in the
computational coordinates, which is challenging for maintaining
accuracy because of the appearance of second angular derivatives of
$\omega$. 

Alternatively, the waveform can be obtained from the asymptotic value
of the Weyl tensor. Asymptotic flatness implies that the Weyl tensor
vanishes at $\scri^+$, i.e. $\hat C_{\mu\nu\rho\sigma}=O(\ell)$ in
the $\hat g_{\mu\nu}$ conformal Bondi frame (\ref{eq:lmet}). This is
the conformal space version of the peeling property of asymptotically
flat spacetimes~\cite{Penrose}. Let $(\hat n^\mu, \hat \ell^\mu, \hat
m^\mu)$ be an orthonormal null tetrad such that $\hat n^\mu=\hat
\nabla^\mu \ell$ and  $\hat \ell^\mu \partial_\mu=\partial_\ell$ at
$\scri^+$. The radiation is described in this frame by the limit 
\begin{equation}
      \hat \Psi:=-\frac{1}{2} \lim_{\ell \rightarrow 0}\frac{1}{\ell}
    \hat n^\mu \hat m^\nu \hat n^\rho \hat m^\sigma \hat C_{\mu\nu\rho\sigma},
\label{eq:psi}
\end{equation}
which in Newman-Penrose notation~\cite{NP} corresponds to
\begin{equation}
     \hat \Psi=-(1/2)\bar \psi_4^0.
\label{eq:psi40}
\end{equation}
The limit is independent of how the tetrad is extended off
$\scri^+$.

A major calculational result in~\cite{strat} is that
\begin{equation}
       \hat \Psi=\frac{1}{2}\hat n^\mu \hat m^\nu \hat m^\rho 
 \bigg( \hat \nabla_\mu  \hat \Sigma_{\nu\rho}  
       -\hat \nabla_\nu \hat \Sigma_{\mu\rho}\bigg )|_{\scri^+} ,
\label{eq:psisigma}
\end{equation}
where $\hat \Sigma_{\alpha\beta}$ is given by (\ref{eq:Sigma})
and where (\ref{eq:psisigma}) is independent of the freedom
\begin{equation}
      \hat m^\nu \rightarrow \hat m^\nu +\lambda \hat n^\nu.	          
\label{eq:mnfreedom}
\end{equation}

In the same inertial polarization frame used in describing the
news,
\begin{equation}
      \Psi=\frac{1}{2} \omega^{-3}e^{-2i\delta}
    \hat n^\mu F^A F^B \bigg( 
        \partial_\mu  \hat \Sigma_{AB}
       -\partial_A \hat \Sigma_{\mu B}
       - \hat \Gamma^\alpha_{\mu B}\hat \Sigma_{A \alpha}
       +\hat \Gamma^\alpha_{A B}\hat \Sigma_{\mu\alpha}
                      \bigg)|_{\scri^+} .
\label{eq:psia}
\end{equation}
An explicit expression for $\Psi$ in terms of the asymptotic metric coefficients
involves lengthy algebra which was carried out using a Maple script to write it
in terms of $\eth$ operators acting on the spin-weighted computational fields
and to construct the final Fortran expression for $\Psi$.

In inertial Bondi coordinates, (\ref{eq:psia}) reduces to the single term
\begin{equation}
         \Psi = \frac {1}{4} Q^A Q^B\partial_u^2  c_{AB} = \partial_u^2
               \partial_l J|_{\scri^+} . 
               \label{eq:psiinert}
\end{equation}
This is related to the expression for the news function in
inertial Bondi coordinates by
\begin{equation}
       \Psi =\partial_u N.
\label{eq:PsiNu}
\end{equation}
However, as in the case of the news, the full expression for $\Psi$ obtained
from  (\ref{eq:psia}) must be used in the code. This introduces additional
challenges to numerical accuracy due to the large number of terms and the
appearance of third angular derivatives.

These difficulties can be appreciated by considering the linearized
approximation, for which considerable simplification arises. To first order in a
perturbation off a Minkowski background,  the nonlinear expression
(\ref{eq:psia}) for $\Psi$  reduces to
\begin{equation}
      \Psi=\frac{1}{2}\partial_u^2 \partial_\ell J -\frac{1}{2}\partial_u J
      -\frac{1}{2}\eth L -\frac{1}{8} \eth^2( \eth \bar L +\bar \eth L)
       + \partial_u \eth^2 H.
\label{eq:linPsi}
\end{equation}
In the same approximation, the news function is given by
\begin{equation}
   N =\frac{1}{2} \partial_u \partial_\ell J
      +\frac{1}{2} \eth^2(\omega +2H).
\label{eq:linN}
\end{equation}
The linearized Einstein equations imply that (\ref{eq:PsiNu}), i.e. $\Psi
=\partial_u N$, still holds in the linearized approximation. (In the nonlinear
case, the derivative along the generators of $\scri^+$ is $\hat n^\nu
\partial_\nu =e^{-2H}(\partial_u +L^A \partial_A)$ and (\ref{eq:PsiNu}) must be
modified accordingly.)

The linearized expressions (\ref{eq:linPsi}) and (\ref{eq:linN}) provide a
starting point to compare the advantages between computing the radiation via the
Weyl component $\Psi$ or the news function $N$. The troublesome  terms involve
$L$, $H$ and $\omega$, which all vanish in inertial Bondi coordinates. One main 
difference is that $\Psi$ contains third order angular derivatives, e.g. the
term $\eth^3 \bar L$, as opposed to second angular derivatives in the case of
$N$. This means that smoothness of the numerical error is more crucial in
the $\Psi$ approach. Balancing this, another main difference is that $N$
contains the term $\eth^2 \omega$, which is a potential source of numerical
error since $\omega$ must be propagated across the stereographic patch
boundaries via (\ref{eq:omegadot}). Test comparisons of waveforms
obtained via $N$ and $\Psi$ are given in the next section.

\section{Tests of modifications to the stereographic grid}
\label{sec:tests}

The characteristic evolution carried out by the PITT code integrates the
Bondi-Sachs equations by means of a finite difference
approximation~\cite{isaac,highp}. Stereographic coordinates $x^A=(q,p)$
are used to label the angles on the outgoing null cones. In the
original code, two square stereographic patches were used, one centered
about the North pole and the other about the South pole. In the new
stereographic scheme introduced in~\cite{strat}, the patches were
modified to have circular boundaries located just past the equator, and
angular dissipation was introduced to suppress the short wavelength
noise introduced by interpatch interpolation. In addition, in the
original code $\eth$-derivatives were approximated by second order
accurate finite difference approximations. In the present version used
in this paper, the $\eth$-derivatives have been increased to fourth
order accuracy. Although the overall second order convergence rate of the
PITT code remains unchanged, these changes are expected to lead to more
accurate waveforms.

There has been extensive testing of the accuracy of past versions of the code
in~\cite{highp,cce,mod,strat}. Here we repeat some of the linear wave tests
presented in~\cite{strat} in order to demonstrate the improvement obtained by
fourth order accurate angular derivatives. First, in order to verify that the new
treatment of stereographic patches is capable of producing a
fourth order accurate evolution, we carry out a test of wave
propagation on the sphere based upon solutions to the 2D wave equation
\begin{equation}
    - \partial_t ^2 \Phi + \eth \bar \eth \Phi = 0,
\end{equation}
where $\Phi =cos(\omega t) Y_{lm}$, $\omega =\sqrt{l(l+1)}$
and $Y_{l m}$ are spherical harmonics. For the case $l=m=2$ we
measure the convergence rate of the error. The simulations are run
with $n+1$ grid points along the axes of each patch, with the grid sizes
ranging from $n=80$ to $n=240$.  For a given grid size, we use the
$L_\infty$ norm to measure the error 
\begin{equation}
       {\cal E}(\Phi) =||\Phi_{numeric}-\Phi_{analytic}||_\infty
\end{equation}
for the circular patches in each hemisphere. We measure the convergence
rate for ${\cal E}(\Phi)$ at a given time $t$, for two consecutive grid
sizes $n_1$ and  $n_2$, by
\begin{equation}
    {\cal R} = \frac{\log_2 \big ({\cal E}(\Phi)_{n_2}
                            / {\cal E}(\Phi)_{n_1} \big )} 
                {\log_2 \big (n_1 / n_2\big )}.
\label{eq:conv}
\end{equation}
Convergence rates for the derivatives are measured analogously.

Excellent 4th order convergence of ${\cal E}(\Phi)$ was obtained. It is
more important and challenging for assessing waveform extraction error 
to measure the error in the derivatives $\eth \Phi$, $\eth^2\Phi$, and
$\eth^3\Phi$, since second angular derivatives enter in the computation
of the Bondi news and third angular derivatives enter into the
computation of $\Psi$. The convergence rates, measured with the
$L_\infty$ norm over the North patch, are shown in
Table~\ref{table:2Dwave} based upon the grid sizes
$(n_1,n_2) = (80,120), (120,160), (160,200), (200,240)$.

\begin{table}[htdp]
\caption{Convergence rates for errors in $\eth \Phi$, $\eth^2\Phi$ and $\eth^3\Phi$}
\begin{center}
\begin{tabular}{|c|cccc|}
	\hline
$ {\cal E}/n$                        &  $n_1=80$  & $n_1=120$  &  $n_1=160$  & $n_1=200$ \\
	\hline
$ {\cal E}(\eth \Phi)   $      &   $4.04$  & $4.11$  &   $4.35$  & $4.85$  \\
$ {\cal E}(\eth^2\Phi)$  &    $4.07$  & $4.24$  &   $4.80$  & $3.95$  \\
$ {\cal E}(\eth^3\Phi)$  &    $3.98$  & $3.95$  &   $3.92$  & $3.86$  \\ [1ex]
	\hline
\end{tabular}
\end{center}
\label{table:2Dwave}
\end{table}%

For the coarser grids, good 4th order convergence is apparent for all
the derivatives. As the grids are refined, the error eventually
approaches (double precision) roundoff error and convergence becomes a
moot question.

Next we compare the accuracy of waveform  extraction by computing the news
function $N$ and the Weyl tensor component $\Psi$ in the test problem considered
in~\cite{strat}, which is based upon a periodic, linearized gravitational wave
on a Minkowski background (see Sec. 4.3 of~\cite{BS-lin}). The linearized wave
is expressed in Bondi-Sachs  coordinates so that it allows direct measurement of
the numerical error. The wave has period $T= \pi$ and $(l=2,m=0)$ spherical
harmonic dependence, with the maximum value of $J\approx10^{-6}$. The data
provided by the linearized solution at the extraction worldtube was propagated
to $\scri^+$ by the characteristic code, where the waveform was computed and
compared to its analytic value.  The computational error in the waveform was
measured with the $L _2$ norm over the North patch using the $n=100$ grid.

\begin{figure}[htp] 
    \centering
    \begin{psfrags}
    \psfrag{xlabel}{t}
    \psfrag{ylabel}{${\cal E}(N)$}
    \includegraphics*[width=12cm]{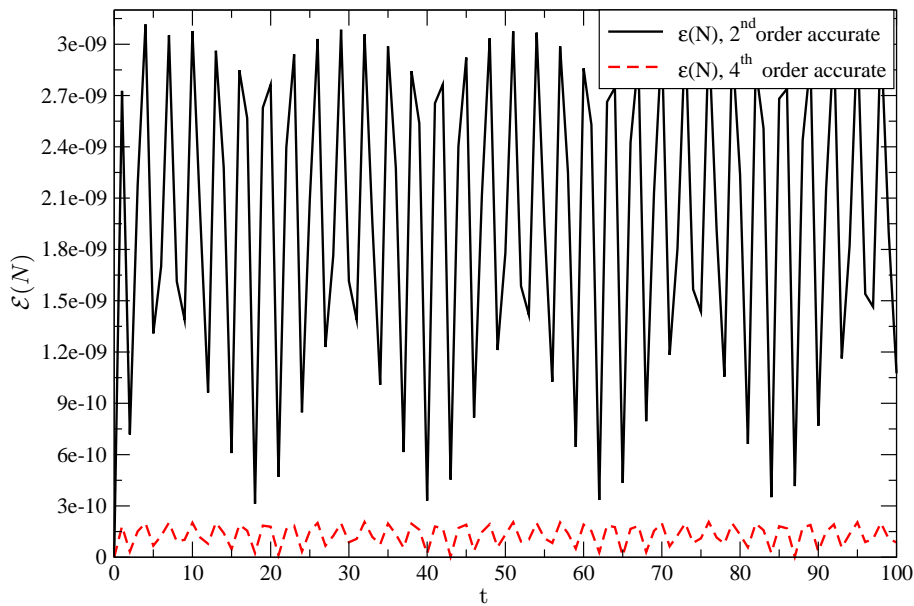}
    \end{psfrags}
    \caption{Plots of the $L_2$ errors ${\cal E}(N)$ 
    vs $t$ in the real part of the news function extracted
    in a linearized gravitational wave
    test. The plots compare the errors obtained
    using the $2^{nd}$ and $4^{th}$ order 
    accurate angular derivatives on an $n=100$ grid. The fourth order method
    reduces the error by an order of magnitude. The time variation of
    the error matches the period of the wave.
                   }
    \label{fig:News2nd4th}
 \end{figure}

Figure~\ref{fig:News2nd4th} compares plots of the error in the real
part of the news function $N$ (computed on the North patch) for the
2nd and 4th order accurate angular derivatives. The plots show roughly
one order of magnitude improvement in accuracy for the $n=100$ grid.
The corresponding plots of the error in the waveform measured by the Weyl 
component $\Psi$ show again roughly one order of magnitude improvement in 
accuracy. Further improvement in accuracy might be obtained by also increasing 
the radial derivatives to fourth order approximations but this could entail
nonlinear complications which could affect the numerical stability of the
evolution algorithm~\cite{luisdis}.

\begin{figure}[htp] 
   \centering
  \begin{psfrags}
  \psfrag{xlabel}{t}
  \psfrag{ylabel}{${\cal E}$}
  \includegraphics*[width=12cm]{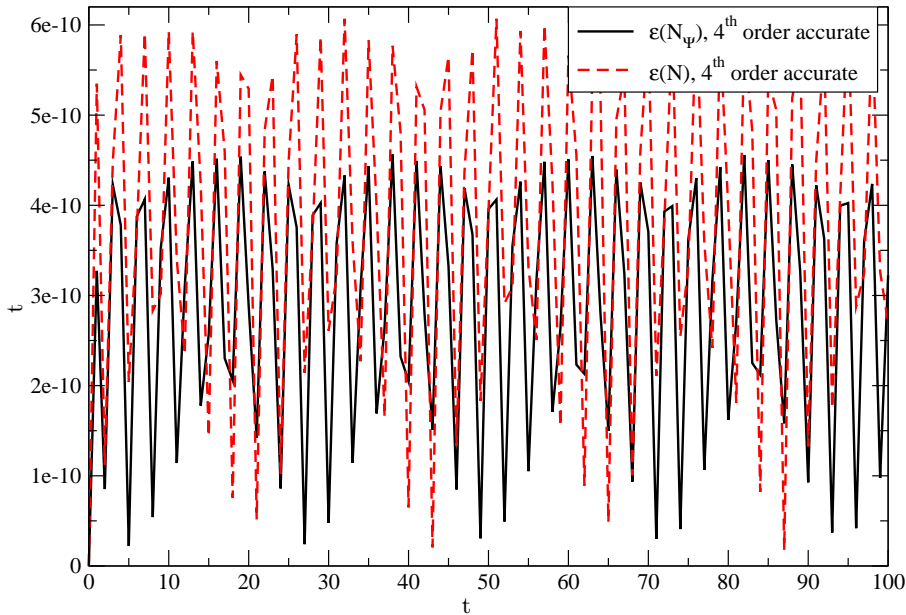}
   \end{psfrags}
  \caption{Comparison plots of the $L_2$ errors  ${\cal E}(N)$ and
  ${\cal E}(N_\Psi)$ vs $t$ in the news function computed directly and
  via the Weyl tensor for the linearized gravitational wave test. The results
  were obtained using $4^{th}$ order accurate angular derivatives on an
  $n=100$ grid. The two methods are competitive although ${\cal E}(N_\Psi)$ 
  is slightly smaller in this case.
               }
   \label{fig:NewsPsi4th}
 \end{figure}

In accord with (\ref{eq:PsiNu}), the  computation of the Weyl component
$\Psi$ yields an alternative numerical value for the news
\begin{equation}
     N_\Psi = N|_{u=0} +\int_0^u \Psi du,
\label{eq:npsi}
\end{equation}
where $N=N_\Psi$ in the analytic problem. Figure~\ref{fig:NewsPsi4th} 
compares these two extraction methods in terms of the errors in $N$ and
$N_\Psi$ for the linearized wave test when using 4th order accurate
angular derivatives. The plots show that the two methods are
competitive although the error in  $N_\Psi$ is slightly smaller in this
case.

\section{Computational interface}
\label{sec:interface}

We have designed an interface that takes Cartesian grid data from a
Cauchy evolution and converts it into boundary data for a characteristic
evolution on a spherical grid extending to $\scri^+$. We treat each
component $g_{\mu\nu}(t,x^i)$ of the Cauchy metric as a scalar function
in the $x^i$ Cartesian coordinates which are used in the $3+1$
evolution. In order to make the interface as flexible as possible for
use as a community tool for waveform extraction, we
have based it upon a spectral decomposition of the Cauchy data in the
region between two world tubes or radii $R=R_1$ and $R=R_2$, where
$R=\sqrt{\delta_{ij}x^i x^j}$ is the Cartesian coordinate radius. Then
at a given time $t$, we decompose $g_{\mu\nu}(t,x^i)$ in terms of
Chebyshev polynomials of the second kind $U_k(R)$ and spherical
harmonics $Y_{l m}(\theta,\phi)$, where $(\theta,\phi)$ are related
to $x^i/R$ in the standard way. The Chebyshev polynomials are
conventionally defined as functions $U_k(\tau)$ on the interval $-1 \le
\tau \le 1$. Here we map them to the interval $R_1 \le R \le R_2$ by
the transformation
$$
      \tau (R)=\frac{2R-R_1 -R_2}{R_2 -R_1} .
$$
Thus, for $R_1<R<R_2$, we expand 
\begin{equation}
    g_{\mu\nu} (t,x^i) = \sum_{kl m} C_{\mu \nu [kl m]}(t)
   U_k (R) Y_{l m}(\theta,\phi).
\end{equation}
For the application to waveform extraction given in this paper, we
choose $l\le l{Max}$, where $l{Max} =6$, and
$k\le k_{Max}$, where $k_{Max} =6$. These values should
be considered tentative and further experimentation
is warranted to optimize accuracy. In tests of binary black holes
with mass $M$ we use a relatively small range $R_2 -R_1 =10M$ and a
larger value of $k_{Max}$ would certainly be needed if the range were expanded.
Also, while $l{Max} =6$ might be sufficient for extraction at $R_E=100M$, a larger
value might give improved results at $R_E=20M$.

The coefficients $C_{\mu \nu [klm]}$ allow us to reconstruct a
spherical harmonic decomposition of each component of the Cauchy metric
on the extraction worldtube $R=R_E$, i.e. 
\begin{equation}
   g_{\mu\nu [l m]}(t,R_E)
         =  \sum_{k} C_{\mu \nu [kl m]}(t) U_k (R_E) .
\end{equation}
This decomposition is carried out at a sequence of Cauchy time steps
$t_n=t_0 +n\Delta t$, where $\Delta t$ is chosen to be much smaller
than the physical time scales in the problem but, for purpose of
economy, larger than the time step used for the Cauchy evolution.
At each time step, the spectral coefficients are determined by a
least squares fit to the Cauchy metric.

The extraction module also requires the derivatives $\partial_t g_{\mu\nu}$ and
$\partial_R g_{\mu\nu}$ at the extraction worldtube. The $R$-derivative is
obtained analytically, at each time level $t_n$, by differentiation of the
Chebyshev polynomials. In one option, the finite difference option, the
$t$-derivative is constructed by a fourth-order accurate finite difference
approximation based upon the sequence of Cauchy times $t=t_n$. In a second
option, the fast-Fourier-transform  option,  we modify the Cauchy data by 
filtering each mode $f_n=C_{k l m} (t_n)$ to  remove high-frequency noise (both
numerical noise and high-frequency gauge waves). The filter works as follows. Let
$f_n$ be the original data $(n=0, \dots ,N-1)$, and
$g_n = f_n - a (n \Delta t) -b (n\ \Delta t)^2.$ The coefficients $a$ and $b$ are
fixed by requiring that $g_N = g_0$  (where $g_N$ is extrapolated from $g_{N-1}$
and $g_{N-2}$) and $g_0 - g_{N-1} = g_2 - g_1$, i.e. the one sided derivatives
taken at $n=0$ agree. This guarantees continuity of $g$ and its first derivative
when periodically extended.  We then perform a fast Fourier transform on $g_i$,
truncate the transform at high frequencies, and perform an inverse Fourier
transform to obtain a filtered $G_n$ and optionally, the inverse transform of
$i \omega\ g_i$ to obtain a smooth time derivative of $G_n$. We then construct
the filtered mode $C_{k l m} (t_n) = G_n  + a (n\Delta t) + b (n\Delta  t)^2$, as
well as its time derivative.

The stereographic coordinates $x^A=(q,p)$ used to label the outgoing null rays
in the Bondi metric are matched to the spherical coordinates $(\theta,\phi)$
induced by the Cartesian Cauchy coordinates on the extraction worldtube by a
standard transformation, using the conventions in~\cite{eth}. The value of the
surface-area coordinate $r$ in the Bondi-Sachs metric is obtained on the
extraction worldtube from the 2-determinant of the Cartesian metric on the
surfaces $t=t_n,R=R_E$. As a result $r_E(t_n,q,p):=r|_{R=R_E}\ne const$ on the
extraction worldtube. In order to deal with this complication, the
transformation from Cartesian coordinates $(t, x^i)$ to Bondi-Sachs coordinates
$(u,r,x^A)$ is carried out via an intermediate Sachs coordinate system
$(u,\lambda,x^a)$~\cite{sachsl} where $\lambda$ is an affine parameter along the
outgoing null rays. The affine freedom allows us to set $\lambda=0$ on the
extraction worldtube. After carrying out the Jacobian transformation from
$(t,x^i)$ to $(u,\lambda,x^A)$, the Cartesian metric and its first derivatives
at the extraction worldtube provide a first order Taylor expansion in $\lambda$
(about $\lambda =0$) of the null metric in Sachs coordinates. The corresponding
Taylor expansion of the metric in Bondi-Sachs coordinates then follows from the
computed values of $r_E$ and $\partial_\lambda r$ at $\lambda=0$, which are
obtained from the 2-determinant of the Cartesian metric~\cite{ccm}.

This allows us to build a grid based upon the characteristic coordinates
$(x,q,p)$, with compactified radial coordinate $x$ given by (\ref{eq:compx}).
The grid values $x_i =x_{i-1} +\Delta x$, $1\le i \le n_x$, are adjusted so that
$x_1 < x_E:= x|_{R=R_E}$ and $x_{n_x}=1$.  The characteristic time levels
$u_n=u_{n-1}+\Delta u$ are chosen to coincide with  the Cauchy times $t_n$ on
the extraction worldtube by choosing $(u-t)|_{R=R_E} =0$.

In the previous version of the extraction module, the first order Taylor
expansion for the Bondi metric was used to fill the gridpoints neighboring the
extraction worldtube and thus initiate the radial integration of the
hypersurface equations (\ref{eq:beta}) - (\ref{eq:ww}). However, the
hypersurface equations require only 6 (real) integration constants, which can be
supplied by their values at $R=R_E$. Using the Taylor expansion to fill the
neighboring gridpoints leads to a potential inconsistency between the Bondi
metric supplied by the Cauchy evolution and the radial derivatives determined by
the characteristic hypersurface and evolution equations. In particular, we have
found that such inconsistencies arising from error in the Cauchy data degrade
the convergence rate of the characteristic extraction module. Because
convergence of the extraction module is an important  test of its reliability,
we proceed here in a different manner which decouples the Cauchy and
characteristic extraction errors.

In the previous version of the extraction module, the Taylor expansions were
also applied to the auxiliary variables $\nu=\bar \eth J$, ${\cal B}=\eth\beta$ and
$k=\eth K$ by applying the $\eth$-operator to the Taylor expansions of the main
variables. This was a complicated process because the $\eth$ operator intrinsic
to the $\lambda=0$ extraction worldtube is not the same as the $\eth$ operator
intrinsic to the $r=const$ Bondi spheres (as they differ by radial derivatives).
In the process, several bugs were introduced in the radial start-up scheme. The
present version of the extraction module streamlines the start-up of the
auxiliary variables by avoiding the use of Taylor expansions.

In this new approach, the hypersurface equations are integrated purely in terms
of the values $\beta_E$, $Q_E$, $U_E$ and $W_E$ of the hypersurface variables on
the extraction worldtube which are supplied by the Cauchy data. A mask is set up
to identify those radial grid points $i \le B$ (referred to as ``$B$ points'')
for which $x_i - x_E \le \Delta x$. These grid points are  ``passive'' points
which do not directly enter in the evolution. Values of the hypersurface
variables are assigned at the first active points $i=B+1$ (referred to as
``$B+1$ points'') in the following manner, assuming that the values $J_E$ and
$J_{B+1}$ of the evolution variable $J$ are known, as well as the values
$\nu_{B+1}$ and $k_{B+1}$ of the auxiliary variables. (We address the latter
assumption below in describing the start-up of the evolution algorithm.) 
Proceeding in the hierarchical order of the hypersurface equations, we first use 
(\ref{eq:beta}) to determine $\beta_{B+1}$ according to
\begin{equation}
         \beta_{B+1} = \beta_E + N_\beta [J](r_{B+1} - r_E).
\end{equation}
Because $N_\beta [J]$ only involves $J$ and $\partial_r J$ it may be evaluated
at the mid-point between $x_E$ and $x_{B+1}$ so that the resulting error in $
\beta_{B+1}$ is $O(\Delta x^3)$. This also determines the auxiliary variable
${\cal B}=\eth \beta$ at the $B+1$ points provided the $B+1$ points on the
neighboring rays have the same grid value $x_i$. However, in the case of an
irregularly shaped extraction worldtube, there can be exceptions where this
neighboring ray is a $B$ point. As a result, in cases where the $B$ points lie
close to the boundary of the masked region they can couple to the B+1 points on
neighboring rays through the $\eth$ operator. For this reason, we also update
$B$ points by the same scheme used for the $B+1$ points. (If a $B$ point is
within a small tolerance of the world-tube, we instead just copy the world-tube
value rather than risk an ill-conditioned algorithm.) In this way, the start-up
value of the auxiliary variable ${\cal B}_{B+1}$ is determined in all cases. 

Next in the hierarchy of hypersurface equations, we determine $Q_{B+1}$ in
similar fashion. However, $N_Q[J,\beta]$ involves ${\cal B}=\eth \beta$ which
cannot be determined on the extraction worldtube from the values of $\beta_E$
(because of the angular variation of $r_E$  discussed above). Consequently, in
order to start up the $Q$-integration we evaluate $N_Q$ at $x_{B+1}$, where
${\cal B}_{B+1}$ is known. This results in an $O(\Delta x^2)$ error in the value
of $Q_{B+1}$. Similar considerations apply to the start-up of the $U$ and $W$
integrations. As a result, the start-up leads to an overall  $O(\Delta x^2)$
error in values at $x_{B+1}$, which is consistent with the global $O(\Delta x^2)$
error resulting from the remaining integration from $x_{B+1}$ to $\scri^+$. This
radial march to $\scri^+$ proceeds in the same way as described
in~\cite{highp,gomezfo,luisdis} to determine all variables on the hypersurface.

\begin{figure}[htp] 
   \centering
    \includegraphics*[width=12cm]{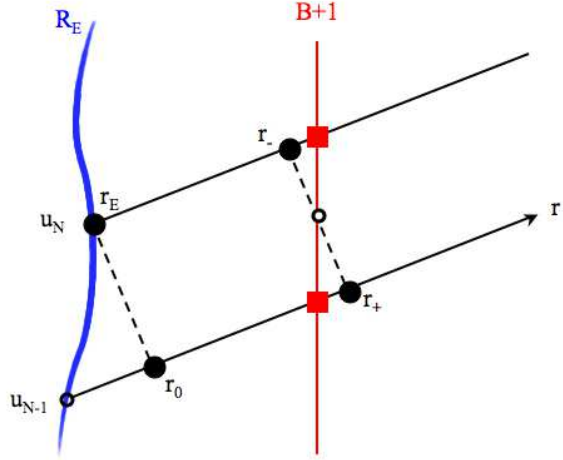}
    \caption{The start-up diagram in the $(u,r)$ subspace for the $B+1$ points
		    (shaded squares). On the left, the extraction worldtube with
		    fixed Cartesian radius $R_E$ moves with respect to the null
		    grid. The null parallelogram for the start-up algorithm is
		    bounded by the two outgoing characteristics at retarded
		    times $u_{n-1}$ and $u_n$ and the two ingoing
		    characteristics indicated by dashed lines. The labels for
		    the radial null  coordinate $r$ are indicated at the four
		    corners (shaded circles). }
    \label{fig:Bp1alg}
 \end{figure}

Having completed the radial march on the hypersurface at time $u_{N-1}$, the
start-up of the integration scheme on $u_n=u_{n-1}+\Delta u$ begins with the
determination of $J_{B+1}(u_n)$ from the worldtube data $J_E$, $\beta_E$, $Q_E$,
$U_E$ and $W_E$ on $u_N$ and the fields  already determined on $u_{N-1}$. We
determine $J_{B+1}(u_n)$ using a null parallelogram algorithm~\cite{nullp}. The
evolution equation (\ref{eq:wev}) for $J$ can be rewritten as
\begin{equation}
             2\partial_u \partial_r \Phi - \partial_r ( A \partial_r \Phi)  = RHS
\end{equation}
where $\Phi=rJ$ and $A=V/r = 1+ rW$. This can be integrated over the null
parallelogram in the $(u,r)$ subspace bounded by the $u_n$ and $u_{n-1}$
hypersurfaces and by two ingoing characteristics. For constant $A$, the ingoing
characteristics satisfy $r - (Au/2) = const$. As depicted in
Fig.~\ref{fig:Bp1alg}, by choosing one ingoing characteristic to pass through
$r_E$ on the $u_n$ hypersurface and the other to pass through $r_{B+1}$ on the
midpoint between the $u_n$ and $u_{n-1}$ hypersurfaces, we obtain the integral
approximation
\begin{equation}
    \Phi(u_{n},r_-) = \Phi(u_{n},r_E)  + \Phi(u_{n-1},r_+) -  \Phi(u_{n-1 },r_0)
                        +\frac { RHS (r_+ -r_0) \Delta u}{2}.
 \label{eq:nullpar}
\end{equation}
Here the corners of the null parallelogram are located at
$r_E$, $r_\pm = r_{B+1} \pm (A\Delta u/4)$ and $r_0=r_E + (A\Delta u/2)$; and the
center of the null parallelogram is located at $r_c = (1/2)(r_E +r_+)$. This
determines the start-up value $\Phi(u_{n},r_{B+1})$ through the second order
accurate interpolation
\begin{equation}
      \Phi(u_{n},r_-)= \frac
    {[\Phi(u_{n},r_{B+1})-\Phi(u_{n},r_E)]r_-}{r_{B+1} -r_E}.
\end{equation}
Using the worldtube data and field values on $u_{n-1}$, all other quantities can
be approximated consistent with second order accuracy  except for a term in
$RHS$ which is proportional to $\partial_u J$. This term is treated to the
required accuracy by a two-step Crank-Nicholson iteration, as is done in the
main evolution scheme described in~\cite{highp}. This leads to a value of
$\Phi(u_{n},r_{B+1})$, and thus $J(u_{n},r_{B+1})$, with $O(\Delta x \Delta
u^2)$ error. As in the case of the hypersurface equation, we also use this
algorithm to update $J$ at the $B$ points to assure that the auxiliary variables
$\nu_{B+1}$ and $k_{B+1}$ can be determined by application of the $\eth$
operator. Now the radial march continues to the $B+2$ points by a similar
process.

\bigskip

\subsection{Convergence measurements}

In tests of the waveform and other variables obtained from a binary black hole
evolution there are no exact values available for measuring error so that
convergence rates cannot be obtained by use of (\ref{eq:conv}). Instead, we
obtain Cauchy convergence rates by using measurements obtained with three
different gridsizes. For grids in the ratio $\Delta_3= \chi \Delta_2 =\chi^2
\Delta_1$ the Cauchy convergence rate of a measured quantity $F$ is given
\begin{equation}
    {\cal R} = \frac{\log_2 \big ((F_3 -F_2)/(F_2-F_1) \big )} 
                {\log_2 \chi}.
\label{eq:cconv}
\end{equation}
For quantities that approach the continuum value $F_0$ as $F=F_0 + G \Delta^2$,
(\ref{eq:cconv}) gives a convergence rate of 2 when $G$ is a smooth function
independent of gridsize. In the main part of the characteristic evolution
algorithm, $G$ is determined by the second derivatives of the evolution
variables. However, a stochastic grid-dependent source of second order error occurs
in the start-up algorithm due to the location of the $B+1$ points. The
separation $x_{B+1}-x_E$ of this point from the extraction worldtube can vary
discontinuously under a small change in gridsize, i.e.
$x_{B+1}-x_E = (1+\epsilon) \Delta x$, where $\epsilon$ is a random number,
$0 <\epsilon <1$. This random
separation enters into the second order accurate approximations made in the
start-up algorithm. The approach to the continuum value has the form $F=F_0 +
(G+\tilde G \epsilon) \Delta^2$. Consequently, the stochastic part of the second
order error can obscure the convergence rate determined by (\ref{eq:cconv}) if
$\tilde G$ is comparable in size to $G$. The only way to ensure clean
convergence rates would be to implement a third order accurate start-up
algorithm, which would involve a considerable amount of work. Fortunately,
this source of error does not appear to be significant in the tests we have
carried out. All the main variables exhibit second order convergence
when measured at a finite radius for the results of the binary black hole inspiral
presented in Sec.~\ref{sec:bbhwaveforms}. However, some asymptotic quantities
at $\scri^+$ display convergence rates intermediate between first and second
order, for reasons discussed further in Sec.~\ref{sec:bbhwaveforms}. 

\subsection{Constraints on the time step}

Domain of dependence considerations place a constraint between the
characteristic time step $\Delta u$ and the size of the characteristic
grid analogous to the CFL condition for the Cauchy evolution. For a rough
estimate, consider the Minkowski space case with the conformally
rescaled metric 
\begin{equation}
    ds^2=-\frac{(1-x)^2}{R_E^2} du^2 -\frac {2}{R_E} du dx
       +  q_{AB} dx^A dx^B
\end{equation}
in the compactified stereographic coordinates $(u,x,q,p)$
used in the code, for which the unit sphere metric takes the form
\begin{equation}
    q_{AB} dx^A dx^B = \frac{4}{1+p^2 +q^2} (dp^2 + dq^2). 
\end{equation}
The past light cone is determined by 
\begin{equation}
    \frac {du}{R_E} = \frac {-dx 
       - \sqrt {dx^2+(1-x)^2 q_{AB} dx^A dx^B}} {(1-x)^2}.
\end{equation}
The restriction on the characteristic time step arising from  domain of
dependence considerations is strongest at the inner boundary, where $x=1/2$
(since $r_E=R_E$ in the Minkowski case); and it is also strongest at the
equator, where $p^2+q^2 =1$. At these points 
\begin{equation}
    \frac {|du|}{4 R_E} = dx + 
        \sqrt{ dx^2+(1/4)(dp^2+dq^2 ) }.
\end{equation}
For typical characteristic grid parameters,
$\Delta p = \Delta q = \Delta x /4$, the resulting restriction is 
\begin{equation}
    \frac {|\Delta u|}{R_E} < 8 K \Delta x 
\label{eq:cfl}
\end{equation}
where $K \approx 1$ depends upon the details of the finite difference
stencil. This restriction is strongest for a small extraction
radius. The characteristic code monitors the corresponding restriction
on $\Delta u$ determined by the curved space version of the
compactified Bond-Sachs metric. For a Cauchy simulation of binary black
holes with mass $M$ with timestep $\Delta t =M/32$ (sufficient to
describe the frequencies typical of a binary system), (\ref{eq:cfl})
leads to
\begin{equation}
    \frac {M}{256 R_E} <  K \Delta x ,
\end{equation}
for the choice of characteristic timestep $\Delta u = \Delta t$. The
corresponding number of radial gridpoints must roughly satisfy $n_x < 128 R_E/M$.
This places no limit of practical concern on the resolution of the characteristic
evolution even for the small extraction radius $R_E =20 M$. Thus, for purposes of
CCE, there are no demanding CFL restrictions on the characteristic grid and
timestep.

\subsection{Initial characteristic data}
\label{sec:initchar}

The initial data for characteristic evolution consist of the values of $J$ on the
hypersurface $u=T_0$. One way of attempting to suppress incoming radiation in this
data is to set the Newman-Penrose Weyl tensor component $\psi_0=0$ on the initial
null hypersurface. For a perturbation of the Schwarzschild metric, this condition
implies no incoming radiation in the linearized approximation. The condition that
$\psi_0$ vanish involves two-radial derivatives of $J$, which in the compactified
coordinate $\ell=1/r$ takes the simple linearized form $\partial_\ell^2 J$ =0.
Translated into the computational coordinate $x=1/(1+R_E \ell)$, we choose the
solution 
\begin{equation}
     J= \frac{J|_{x_E} (1-x)x_E}{ (1-x_E)x}, 
\end{equation}
which provides continuity of $J$ with its value determined by Cauchy data at the
extraction worldtube. Since this choice of $J$ also vanishes at infinity, the
initial slice of $\scri^+$ has unit sphere geometry and equation (\ref{eq:conf})
for the conformal factor has the simple solution $\omega=1$.

\section{Binary black hole measurements and waveforms}
\label{sec:bbhwaveforms}

Here we present test results of waveform extraction from the inspiral and  merger
of two equal-mass, non-spinning black holes.  For the Cauchy evolution we used the LazEv
code~\cite{Campanelli:2005dd,Zlochower:2005bj} along with the Cactus
framework~\cite{cactus_web} and Carpet~\cite{Schnetter:2003rb} mesh refinement
driver. LazEV is a 8th order accurate finite difference code based upon the
Baumgarte-Shapiro-Shibata-Nakamura (BSSN) formulation~\cite{bssn1,bssn2} of
Einstein's equations, which deals with the internal singularities by the moving
puncture approach~\cite{Campanelli:2005dd, Baker:2005vv}. Our simulation  used 9
levels of refinement with finest resolution of $h= M/80.64$, and outer Cauchy
boundary at $400M$. The initial data consisted of a close quasicircular black-hole
binary with orbital frequency $M\Omega = 0.050$, leading to more than a complete
orbit before merger (See~\cite{Campanelli:2006uy}). We output the metric data on
the extraction worldtube every $\Delta t = M/20$.

In the Cauchy evolution, we extract $\psi_4$ on spheres of Cartesian radius 
$R/M=50, 60, \cdots, 100$ and decompose in spin-weighted $(l,m)$ spherical
harmonic modes. We use a perturbative formula~\cite{Lousto:2010qx} to extrapolate
the perturbative waveform $R \psi_4$ to $R=\infty$,
\begin{equation}
  \lim_{R\to\infty}[R \psi^{l m}_4(R,t) ] = r \psi^{l m}_4(r,t) -
  \frac{(l-1)(l+2)}{2}\int_0^t dt \psi_4^{l m}(r,\tau) d\tau + {\cal O}(r^2),
  \label{eq:extrap}
\end{equation}
where $r$ is the areal radius corresponding to the Cauchy extraction radius R. The
extrapolation of the perturbative waveform to infinity removes cumulative phase
error which otherwise would be introduced by redshift effects. 

We present results for the characteristic extracted waveform either in terms of
$\Psi$, related to the Bondi news by $\Psi=\partial_u N$, or, when comparing to
the perturbative waveform, in terms of the Newman-Penrose component $\Psi_4=-2\bar
\Psi$. For illustrative purposes, we concentrate on the dominant $(l=2,m=2)$ and
sub-dominant $(l=4,m=4)$ spherical harmonic modes.

The highest resolution black hole waveform extraction test was run with the
following characteristic grid specifications: angular gridpoints $n_q=n_p=200$,
radial gridpoints $n_x=224$. For convergence tests, we also used grids $n_q
=n_p=100$, $n_x=112$ and  $n_q =n_p=50$, $n_x=56$, so that the grid sizes were in
the ratio $\chi = 2$. We refer to these as the $n=200$, 100 and 50 grids,
respectively, The characteristic time steps used for these grids were $\Delta t =
M/20  \,\, (n=200)$, $\Delta t = M/10 \,\, (n=100)$ and $\Delta t = M/5  \,\, 
(n=50)$. The characteristic extraction was carried out using worldtube radii
$R_E=20M$, $50M$ and $100 M$. 

The Pitt null code was run on stereographic patches with circular boundaries using
the auxiliary variables (\ref{eq:aux}) to eliminate any second derivatives in the
angular directions and using 4th order accurate angular derivatives. Angular
dissipation was added with the coefficients $\epsilon_x=10^{-3}$,
$\epsilon_u=10^{-4}$, $\epsilon_Q=\epsilon_W=10^{-6}$, in the notation
of~\cite{strat}.

\begin{figure}[htp] 
   \centering
  \begin{psfrags}
  \psfrag{xlabel}{t}
  \psfrag{ylabel}{${\Psi}$}
  \includegraphics*[width=12cm]{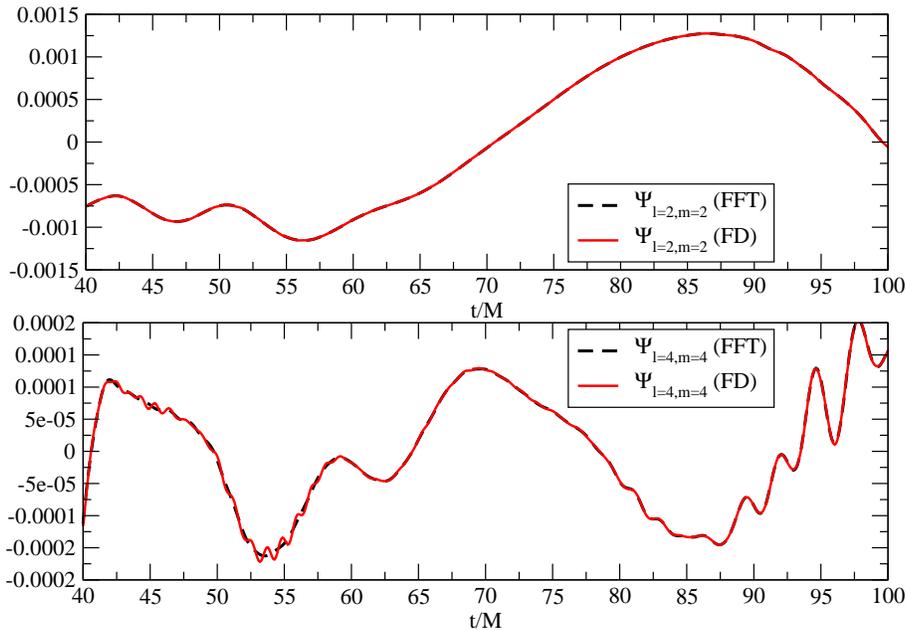}
   \end{psfrags}
  \caption{Comparison of the FD and FFT options for the  $(2,2)$ (up) 
           and $(4,4)$ (down) spherical harmonic modes of the
           real part of the characteristic waveform
           $\Psi$ obtained with the $n=200$ grid and extraction radius $R_E=20M$.
           The two options give comparable results for the $(2,2)$ mode but for the
           $(4,4)$, which is an order of magnitude smaller, high frequency error in the
           world tube data is noticeable in the waveform extracted with the FD option.
               }
   \label{fig:Psi4FFTFD}
 \end{figure}

The best accuracy was obtained using the fast Fourier transform (FFT) option  to
obtain time derivatives of the worldtube data, as described in
Sec.~\ref{sec:interface}. For strong signals, e.g. the dominant  $(l=2,m=2)$
spherical harmonic mode, the finite difference (FD) and FFT options are in good
agreement. However, for weak signals, e.g. the $(l=4,m=4)$ mode, the FD option can
generate noticeable high frequency error. See Fig.~\ref{fig:Psi4FFTFD} for a
comparison of waveforms computed with the two options. One likely source of the
high frequency error with the FD option is the stochastic error introduced at each
time level $t_n$ on the extraction worldtube by the least squares fit of the
Cauchy data to the spectral expansion. In the FD option, this error is amplified
when taking the time derivatives necessary to compute the waveform and becomes
more prominent for short characteristic timesteps. It also becomes more prominent
as the extraction radius is increased, in which case the extracted worldtube data
is smaller. Similar high frequency noise is apparent in the worldtube data so that
this error cannot be removed by refining the characteristic grid. However, the
filtering intrinsic to the FFT option is effective in reducing this error. The
remaining results reported in this paper were obtained with the FFT option.

Convergence rates were measured for the $(l=2,m=2)$ spherical harmonic mode, which
is the dominant mode in the waveform. Table~\ref{bbhconv1} gives the rates for the
evolution variables measured on a sphere at Bondi radius $r=80M$ obtained with a
small extraction radius $R_E=20M$ at a time corresponding to the peak of the
signal ($t\approx 200M$). The rates are given for the real and imaginary part of
the variables. All quantities are very close to second order convergent, including
$J_{,x}$, which is the term which determines the waveform after transforming to
inertial Bondi coordinates according to (\ref{eq:psiinert}).

\begin{table}[htdp]
\caption{Convergence rates of the $(l=2,m=2)$ spherical harmonic mode on the
sphere $r=80M$ for the metric variables measured at retarded time $u\approx 200M$
near the peak of the signal. The rates are given for the real and imaginary part
of the variables. The extraction radius was $R=20M$. The results show that the
evolution variables all display clean second order convergence.} 
\begin{center}
\begin{tabular}{|c|c|c|}
   \hline
   $Variable$ & $ Rate_{Re} $  & $ Rate_{Im} $\\
   \hline
   $\beta$	&	$2.01$	&	$2.01$	\\
   $J$	&	$2.23$	&	$2.01$	\\
   $J_{,x}$	&	$2.03$	&	$2.33$	\\
   $Q$	&	$2.02$	&	$2.04$	\\
   $U$	&	$1.99$	&	$1.96$	\\
   $W$	&	$1.97$	&	$2.00$	\\
     \hline 
\end{tabular}
\end{center}
\label{bbhconv1}
\end{table}%

\begin{figure}[htp] 
   \centering
  \begin{psfrags}
  \psfrag{xlabel}{t}
  \psfrag{ylabel}{$Q$ and $W$}
  \includegraphics*[width=12cm]{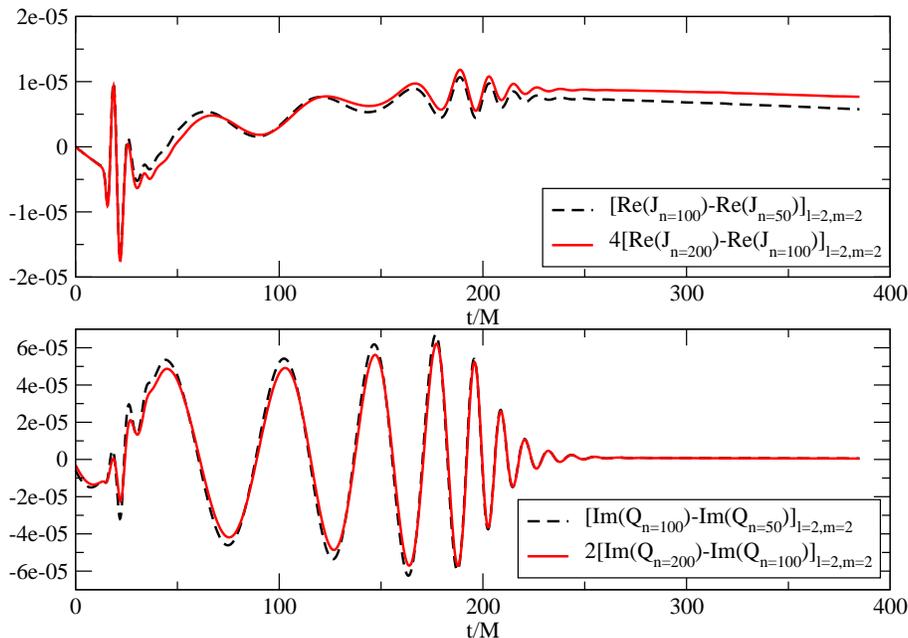}
   \end{psfrags}
  \caption{Convergence plots of the asymptotic limits at $\scri^+$ of the $(2,2)$
	       spherical harmonic modes of $Re J$ and $Im Q$  obtained with
	       resolutions $n=50$, $n=100$ and $n=200$ with an  extraction radius
	       $R_E=20M$. The plots for $Re J$ are rescaled for 2nd order
	       convergence (upper plot),  while the plots for $Im Q$ (lower plot)
	       are rescaled for 1st order convergence. The rescaled differences
	       show that convergence rates at the peak of the signal given in
	       Table~\ref{bbhconv2} are representative of the rates over the
	       entire run.  }
   \label{fig:ImJQConv}
 \end{figure}

Table~\ref{bbhconv2} gives the corresponding convergence rates for these evolution
variables measured at $\scri^+$, again at the time corresponding to the peak of
the signal and with extraction radius $R_E=20M$. In this case $Q$ and $W$ show
deviation from second order convergence, consistent with the asymptotic  error
analysis presented in Sec.~\ref{sec:chform} in relation to (\ref{eq:asymqw}). We
also see that the derivative $J_{,x}$ deviates from second order convergence,
which indicates a need for more accurate finite difference approximations near
$\scri^+$. There are several places in the present code where one-sided difference
approximations are used for derivatives at $\scri^+$. These convergence rates at
the peak of the signal are representative of the rates over the entire run. This
is illustrated in Fig.~\ref{fig:ImJQConv} which plots the rescaled errors of $Re
J$ and $Im Q$ versus time at $\scri^+$.

\begin{table}[htdp]
\caption{Convergence rates of the $(l=2,m=2)$ mode for the metric variables
measured near the peak of the signal at $\scri^+$, with an extraction radius
$R=20M$. As expected from the analysis in Sec.~\ref{sec:chform}, some asymptotic
quantities only display first order convergence.}
\begin{center}
\begin{tabular}{|c|c|c|}
   \hline
   $Variable$ & $ Rate_{Re}$  & $ Rate_{Im} $   \\
   \hline
   $\beta$  	     &	$2.01$	&	$2.01$	\\
   $J$	     &	$1.80$	&	$2.18$	\\
   $J_{,x}$	     &	$1.23$	&	$1.20$	\\
   $Q$	     &	$1.33$	&	$1.19$	\\
   $U$	     &	$1.99$	&	$1.96$	\\
   $W$	     &	$1.55$	&	$1.50$	\\
      \hline
\end{tabular}
\end{center}
\label{bbhconv2}
\end{table}%

Table~\ref{bbhconv3} gives the corresponding convergence rates for the waveform as
measured by the Bondi news $N$ and the Weyl component $\Psi$, again at a time
corresponding to the peak of the signal and with extraction radius $R_E=20M$. We
also show the convergence rate of the inertial time derivative $\partial_u N$
calculated directly from finite differencing $N$. All show roughly first order
convergence. The rate for $\partial_u N$ is slightly better than that for $\Psi$,
although  $\Psi =\partial_u N$ in the continuum limit. The convergence rates of
these quantities are affected by two chief factors: (i) the large number of terms
involved in their calculation and (ii) their dependence on radial derivatives of
the evolution quantities at  $\scri^+$. In all cases, one-sided approximations are
used in several places to compute these radial derivatives. This is already
apparent in the convergence rate for $J_{,x}$ shown in Table~\ref{bbhconv2}.

\begin{table}[htdp]
\caption{Convergence rates of the $(2,2)$ spherical harmonic
mode for the Bondi news $N$, $\partial_u N$ obtained by finite difference,
and the Weyl component $\Psi$, all measured near the peak of the signal
with an extraction radius $R_E=20M$.}
\begin{center}
\begin{tabular}{|c|c|c|}
   \hline
   $Variable$ & $ Rate_{Re} $	& 	$Rate_{Im}$\\
   \hline 
   $N$	     		&	$1.59$		&		$1.56$\\
   $\partial_u N$	&	$1.57$		&		$1.55$\\
   $\Psi$    			&	$1.16$		&		$1.14$\\
      \hline
\end{tabular}
\end{center}
\label{bbhconv3}
\end{table}%

 \bigskip

\begin{figure}[htp] 
    \centering
    \includegraphics*[width=12cm]{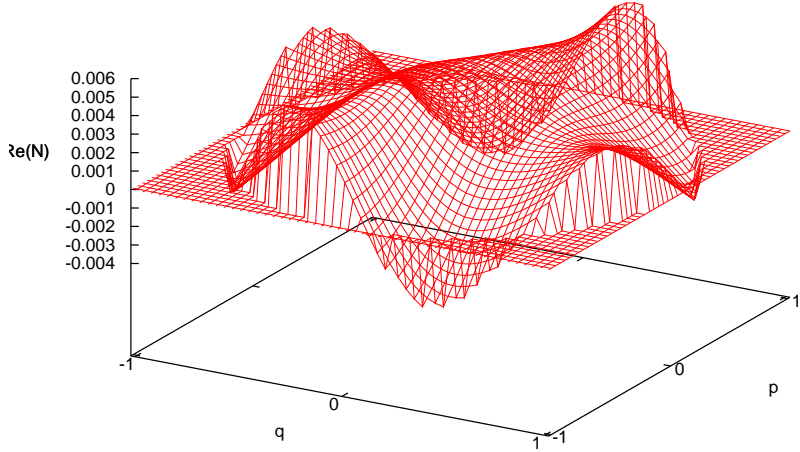}
    \caption{Surface plot in the North $(q,p)$ stereographic patch of the real
		   part of the Bondi news $N$ measured at the peak of the wave
		   with an extraction radius $R_E=20M$. The equatorial patch
		   boundary corresponds to the circle $p^2+q^2=1$. The smooth
		   angular dependence near the equator shows that angular
		   dissipation is effective at removing short wavelength noise
		   arising from interpatch interpolation. } 
    \label{fig:NewsB_scri}
 \end{figure}

\begin{figure}[htp] 
    \centering
    \includegraphics*[width=12cm]{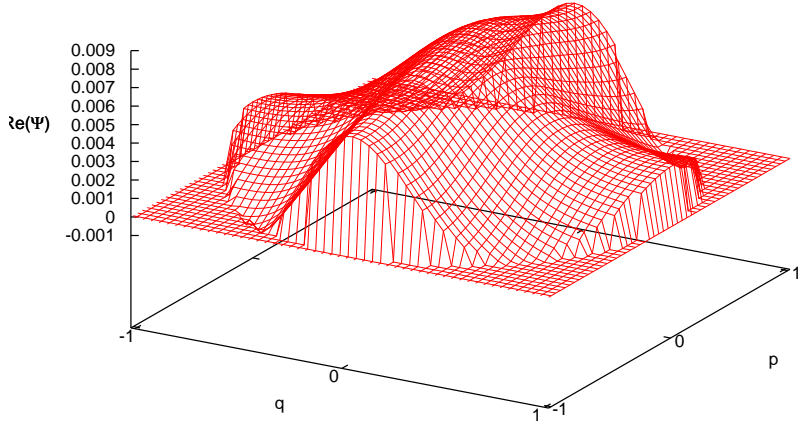}
    \caption{Surface plot in the North $(q,p)$ stereographic patch
                    of the real part of the Weyl component $\Psi$ measured
                     at the peak of the wave with an extraction radius $R_E=20M$.
                     The smooth angular dependence
                     near the equator $p^2+q^2=1$
                     shows that angular dissipation is effective at removing short
                     wavelength noise arising from interpatch interpolation.  } 
    \label{fig:Psi4_scri}
 \end{figure}

Surface plots of the Bondi news $N$ and Weyl component $\Psi$, measured near the
peak of the signal with an extraction radius $R_E=20M$, are shown in
Fig's~\ref{fig:NewsB_scri} and \ref{fig:Psi4_scri}. Both figures display smooth
angular dependence, showing that the angular dissipation is effective at removing
short wavelength angular noise. In particular, there are no ``spikes'' near the
equatorial patch boundary arising from interpatch interpolation. The main error in
the waveform originates from intrinsically time dependent error in the data on the
extraction worldtube. 

\begin{figure}[htp] 
    \centering
    \includegraphics*[width=12cm]{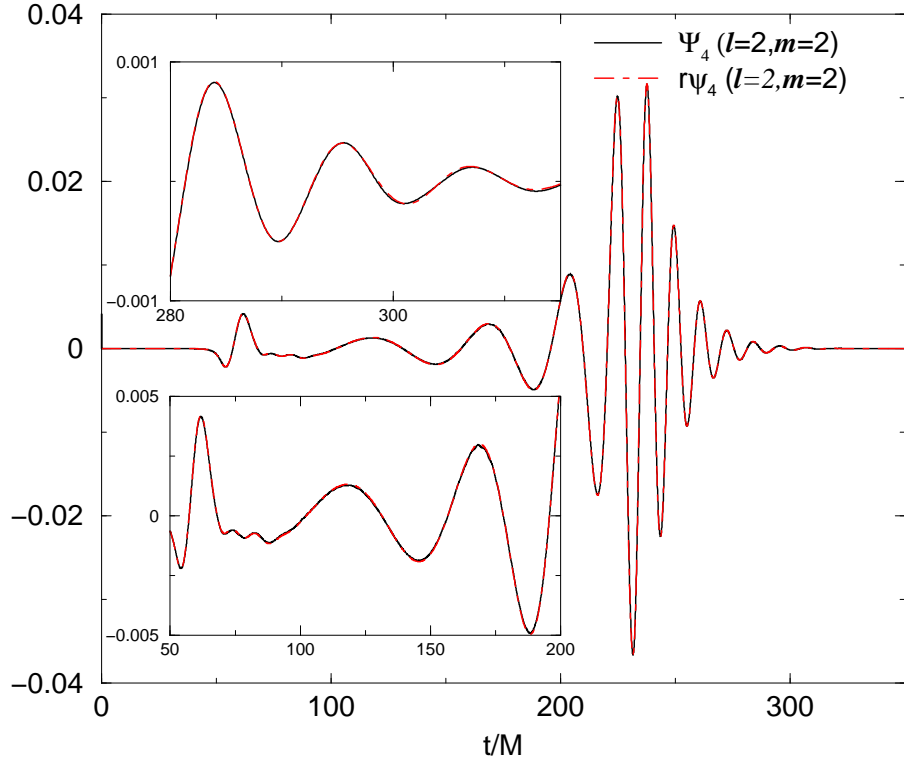}
    \caption{Comparison of the $(2,2)$ dominant spherical harmonic mode for
    $\Psi_4$ (characteristic)  and $r \psi_4$ (perturbative Cauchy), both
    extracted at $R=50M$. The insets show that the excellent agreement extends to
    the early stages and the final ringdown when the amplitude is small.} 
    \label{fig:Psi4_Cc2}
 \end{figure}

\begin{figure}[htp] 
    \centering
    \includegraphics*[width=12cm]{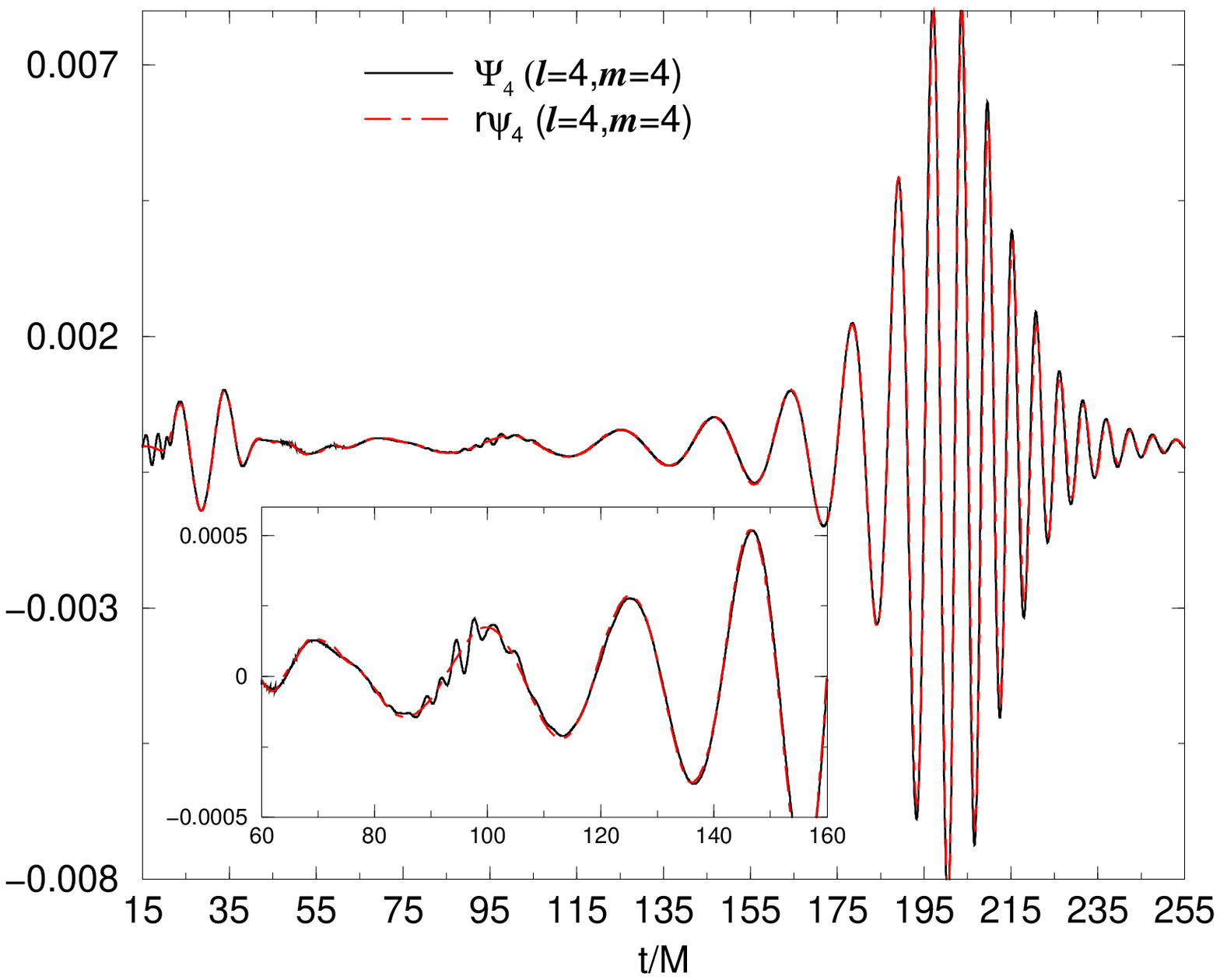}
    \caption{Comparison of the $(4,4)$ sub-dominant mode for $\Psi_4$
     (characteristic) extracted at $R_E=20M$ and $r \psi_4$ (perturbative Cauchy)
     extracted at $R=50M$. There is good agreement in the strong amplitude regime
     of the wave $t/M>120$. At early times $t/M \approx 15$, the characteristic
     waveform exhibits effects of ``junk'' radiation in the initial data near
     $R_E=20M$. In addition, the insert magnifies an anomalous feature, which is
     not fully understood, in the interval about $t/M=90$. } 
    \label{fig:Psi4_Cc4}
 \end{figure}

The time dependence of the real part of the characteristic extracted waveform and
its comparison to the perturbative waveform are shown in Fig's~\ref{fig:Psi4_Cc2}
and \ref{fig:Psi4_Cc4}. Figure \ref{fig:Psi4_Cc2} shows excellent agreement
between these waveforms for the dominant $(l=2,m=2)$ mode, when both are extracted
at $R=50M$. The insets show that this agreement exists in the early stages, when
the amplitude is small, and persists throughout the final ringdown. Note that the
perturbative extrapolation formula (\ref{eq:extrap}) is essential to obtain this
excellent phase agreement between the perturbative and characteristic waveforms.

Figure \ref{fig:Psi4_Cc4} compares the characteristic and perturbative waveforms
for the $(l=4,m=4)$ mode. In this case, the perturbative waveform is again
extracted at $R=50M$ but the characteristic waveform is extracted at $20M$ to
reduce the high frequency noise discussed previously. This high frequency noise
can also be reduced by choosing a larger timestep for the characteristic
evolution, again indicating that it arises from time derivatives of the stochastic
error introduced in the worldtube data by the least squares fit. The
characteristic and perturbative waveforms again show excellent agreement At very
early times $t/M \approx 15$, the characteristic waveform shows an  anomalous
feature which can be attributed to ``junk'' radiation content in the initial data
in the vicinity of $R_E=20M$. In addition, as shown in the insert, there is
another anomalous feature, which is not understood, in the time interval about
$t/M=90$. This feature is also evident in the extracted Cauchy data at $R_E=20M$.
Possible sources for this feature are gauge modes excited in the interior region
or numerical effects from the adaptive mesh refinement used in the Cauchy
evolution. A better explanation would require further runs.

\section{Richardson extrapolation and convergence of the waveform}
\label{sec:rich}

The clean first order convergence results for the news $N$ and Weyl component
$\Psi$ allows us to apply Richardson extrapolation to obtain higher order accuracy
waveforms. We apply the results from the three different resolutions
$n=(50,100,200)$, with grid spacing $(4\Delta, 2\Delta, \Delta)$ respectively, to
obtain a third order accurate waveform as follows. 

The truncation error in a quantity $F$ can be represented by a power series
$$
        F(\Delta)=F_0+F' \Delta + \frac{1}{2}F'' \Delta^2 +O(\Delta^3) .
 $$    
We  write $F_1=f(\Delta)$,   $F_2=F(2\Delta)$   $F_4=F(4\Delta)$.
Then the extrapolated value
 $$
    F_E = \frac{8}{3} F_1-2F_2 + \frac{1}{3} F_4
 $$   
 is 3rd order accurate, i.e.
  $$
    F_E =  F_0 + +O(\Delta^3) .
 $$  
 
In practice this can be confirmed as follows. Let  $ F_I = 2F_1-F_2$ and
$F_{II}=2F_2 -F_4$ be the second order accurate waveforms obtained using data
from two resolutions. Then $ F_{II} -F_E = 4(F_{I}-F_E) +O(\Delta^3)$,  i.e. 
\begin{equation}
     \frac{1}{4}( F_{II} -F_E) = F_{I}-F_E
      \label{eq:check}
\end{equation}
if we neglect the $O(\Delta^3)$ error, i.e. if we approximate the exact value
$F_0$ by the third order accurate approximation $F_E$. 
 
\begin{figure}[htp] 
    \centering
    \includegraphics*[width=12cm]{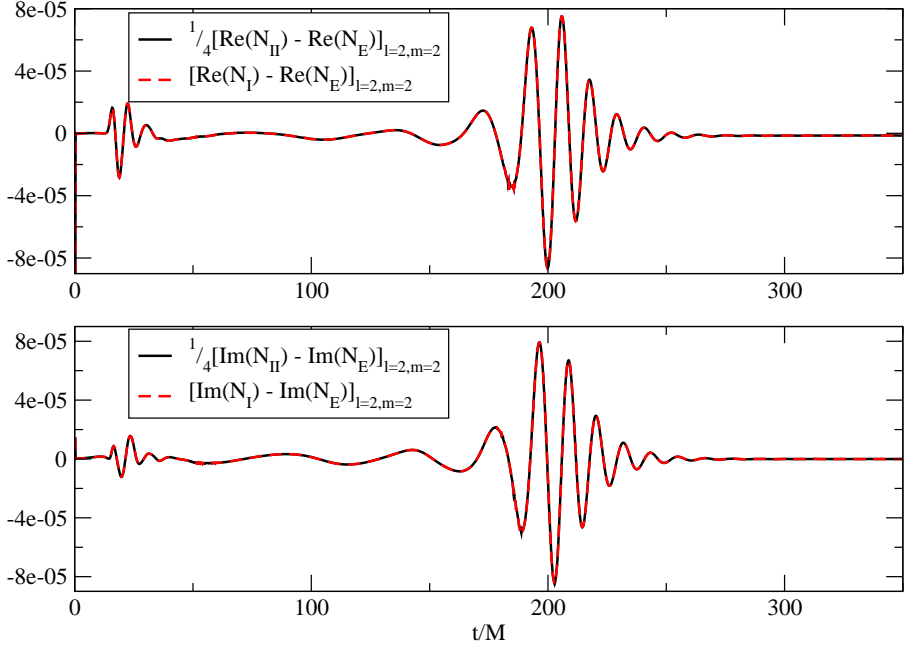}
    \caption{Plots confirming the validity of Richardson extrapolation  to obtain
    higher order accuracy for the real and imaginary parts of the dominant $(2,2)$
    spherical harmonic mode of the news $N(t)$. The rescaled errors show that
    $N_{I}$ and $N_{II}$ are second order accurate in accord with
    (\ref{eq:check}). }
    \label{fig:Ncheck}
 \end{figure}

\begin{figure}[htp] 
    \centering
    \includegraphics*[width=12cm]{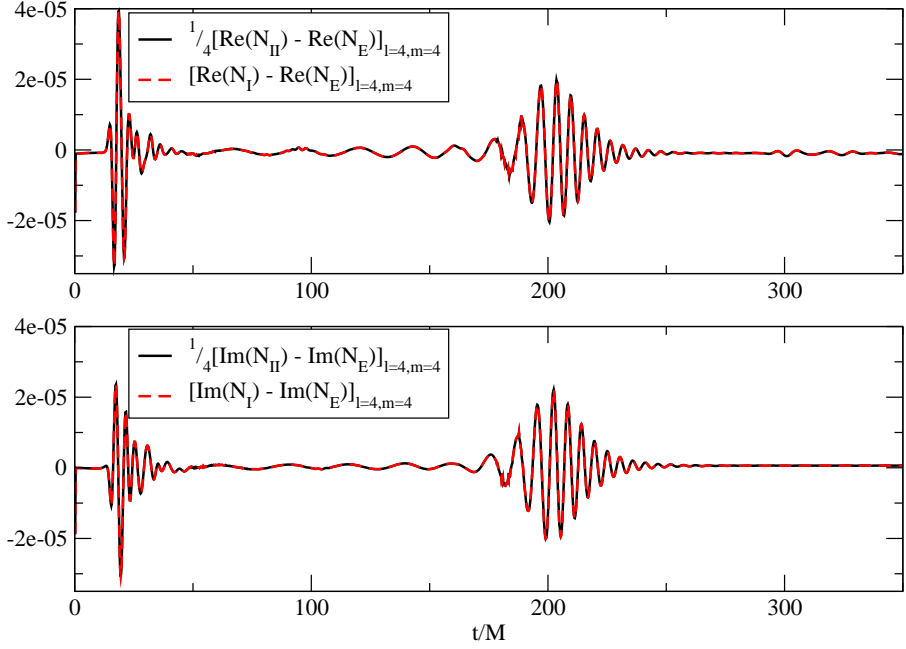} \caption{Plots confirming the validity
    of Richardson extrapolation to obtain higher order accuracy for the
    sub-dominant $(4,4)$  spherical harmonic mode of the news $N(t)$. The rescaled
    errors show again that $N_{I}$ and $N_{II}$ are second order accurate in
    accord with (\ref{eq:check}). }
    \label{fig:Ncheck44}
 \end{figure}

\begin{figure}[htp] 
    \centering
    \includegraphics*[width=12cm]{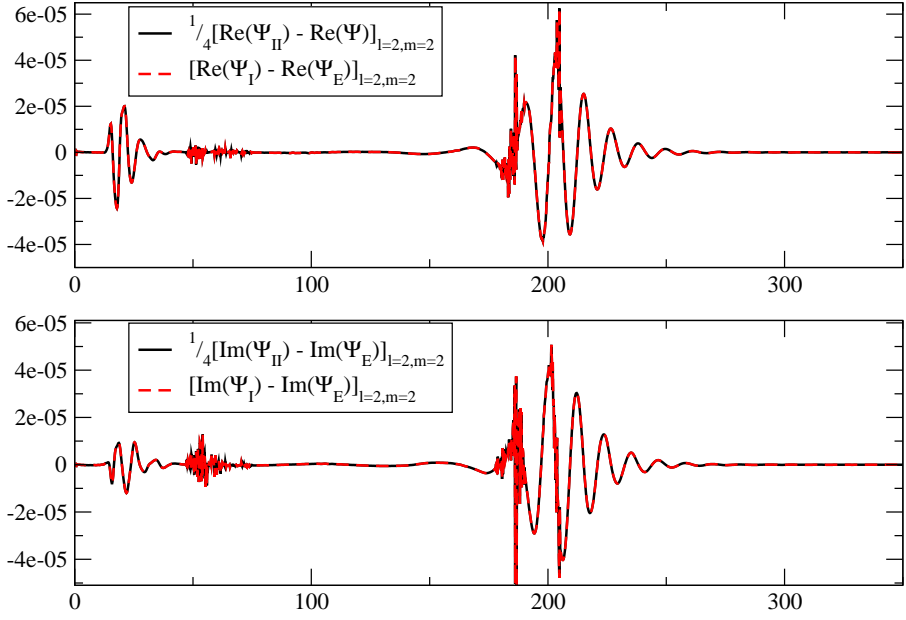}
    \caption{Plots confirming the validity of Richardson extrapolation  to obtain
    higher order accuracy for the real and imaginary parts of the $(2,2)$
    spherical harmonic mode of the waveform $\Psi(t)$. The rescaled errors show
    that $\Psi_{I}$ and $\Psi_{II}$ are second order accurate in accord with
    (\ref{eq:check}). Note that the second order error in $\Psi$ contains more
    high frequency noise than  $N$ shown in Fig.~\ref{fig:Ncheck}. }
    \label{fig:Psicheck}
 \end{figure}

Using the corresponding notation $(N_E, N_I, N_{II})$ for the news and $(\Psi_E,
\Psi_I, \Psi_{II})$ for the Weyl component, we can check the validity of applying 
Richardson extrapolation to the waveform. Figure~\ref{fig:Ncheck} and
Figure~\ref{fig:Ncheck44} graph the rescaled errors of the real and imaginary
parts of  $N_I(t) -N_E(t)$ and $\frac{1}{4}(N_{II}(t)-N_E(t))$ and
Figure~\ref{fig:Psicheck} graphs the corresponding rescaled errors in $\Psi(t)$.
In both cases, (\ref{eq:check}) is confirmed.

These results validate the use of Richardson extrapolation to obtain third order
accurate waveforms $N_E$ and $\Psi_E$. We can use $N_E$ and $\Psi_E$ as fiducial
exact values  and estimate the truncation error in the numerical waveforms by
comparing them with the second order accurate approximates $N_I$ and $\Psi_I$. 
Thus the truncation errors in the news $N$and Weyl  component $\Psi$ are
conservatively given by
 \begin{equation}
     \delta N = N_{I}-N_E =O(\Delta^2)
      \label{eq:Nerr}
 \end{equation}
and 
 \begin{equation}
     \delta \Psi = \Psi_{I}-\Psi_E = O(\Delta^2).
      \label{eq:Psierr}
 \end{equation}
 
\begin{figure}[htp] 
    \centering
    \includegraphics*[width=12cm]{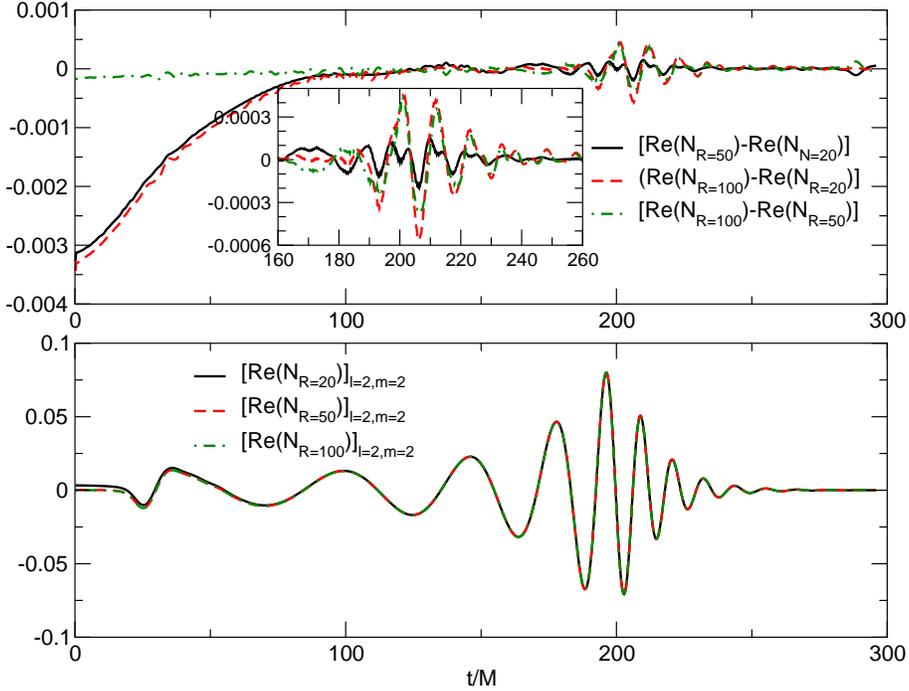}
    \caption{Plots of the $(2,2)$ mode of $Re N$ obtained for extraction radii 
 $R_E=20M$, $50M$, and $100M$. The  $R_E=50M$ and  $R_E=100M$ waveforms have been
 shifted backward in time so that they are in phase at the peak of the wave. The
 noticeable difference in the interval from $t/M=0$ to $t/M=100$ between the
 $R_E=20M$ waveform and the other two results from a mismatch between the initial
 characteristic and Cauchy data, which decreases with large extraction radii. For
 the waveforms extracted at all three radii, the double hump in the interval from
 $t/M=0$ to $t/M=100$  results from non-trivial ``junk'' radiation in the initial
 Cauchy data. The three waveforms are in good agreement in the inspiral and merger
 stage. At the peak of the wave, the relative difference between the $R_E=20M$ and
 $R_E=100M$ waveforms is less than 0.6\% }
    \label{fig:NewsR20R50R100}
 \end{figure}

Figure~\ref{fig:NewsR20R50R100} plots the differences between the dominant
$(l=2,m=2)$ mode of the Richardson extrapolated waveform $N_E(t)$ obtained with
extraction radii  $R_E=20M$, $R_E=50M$ and $R_E=100M$. In the plot,  the $R_E=20M$
waveform begins at $t=0$ and the other waveforms have been shifted backwards in
time so that all three are in phase at the peak of the wave. Two sources of
extraneous ``junk'' radiation can be seen in the figure. One arises from a {\em
mismatch} between the initial characteristic and Cauchy data. The initial
characteristic data $\psi_0 =0$ (see Sec.~\ref{sec:initchar}) implies the absence
of initial radiation content on the assumption that the geometry of the initial
null hypersurface is close to Schwarzschild. This assumption becomes valid as the
extraction radius becomes large and the exterior Cauchy data can be approximated
by Schwarzschild data. Thus this mismatch is largest for extraction at $R_E=20M$.
This results in a noticeable difference at very early times between extraction at
$R_E=20M$ and the other two radii. After $t/M \approx 100$, the three waveforms
are in good agreement  with their relative differences less than 0.6\% at the peak
of the wave.

The second source of  ``junk'' radiation apparent in
Figure~\ref{fig:NewsR20R50R100} arises from the choice of conformally flat initial
Cauchy data.  This arises for all three extraction radii and accounts for the
double hump  in the news function in the interval from $t/M=0$ to $t/M =50$.

\begin{figure}[htp] 
   \centering
   \includegraphics*[width=12cm]{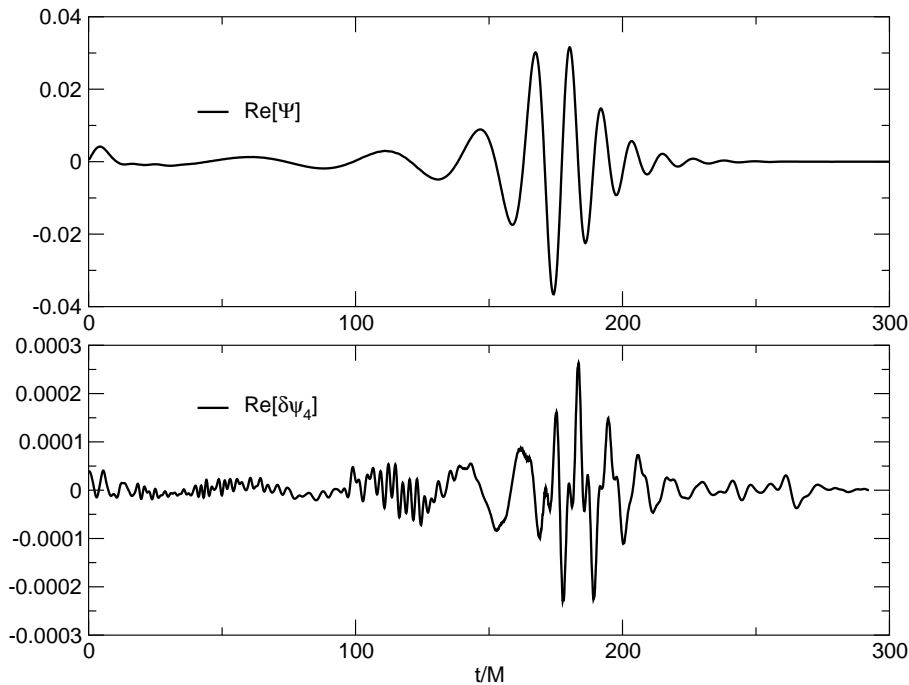}
   \caption{Plots of the time dependence of the real part of the $(2,2)$ spherical
   harmonic components of $\delta \psi_4(t)$, as defined in (\ref{eq:psi4err}),
   and the characteristic waveform $\Psi(t)$. Here $\delta \psi_4(t)$ measures the
   difference between the perturbative and characteristic values of $\Psi(t)$. The
   approximate 1\%  ratio between the peak amplitudes of $\delta \psi_4(t)$ and
   $\Psi$ gives an estimate of the difference between perturbative and
   characteristic extraction. }
   \label{fig:magdiff}
 \end{figure}

It is of interest to measure the difference
\begin{equation}
         \delta \psi_4 = (\frac{1}{2} r \psi_4 +\bar \Psi)
         \label{eq:psi4err}
\end{equation}         
between the extracted waveform using the perturbative extrapolation
formula (\ref{eq:extrap}) and the Richardson extrapolated characteristic
waveform, measured in accord with the normalization conventions indicated in
(\ref{eq:psi40}). Figure~\ref{fig:magdiff}  plots the real part of the $(2,2)$
spherical harmonic component of $ \delta \psi_4(t)$, compared with the
corresponding component of $Re \Psi$. The peak amplitude of $ \delta \psi_4(t)$ is
approximately 1\% of the peak amplitude of $\Psi$, which provides an estimate of
the difference between perturbative and characteristic extraction. 

\begin{figure}[htp] 
    \centering
    \includegraphics*[width=12cm]{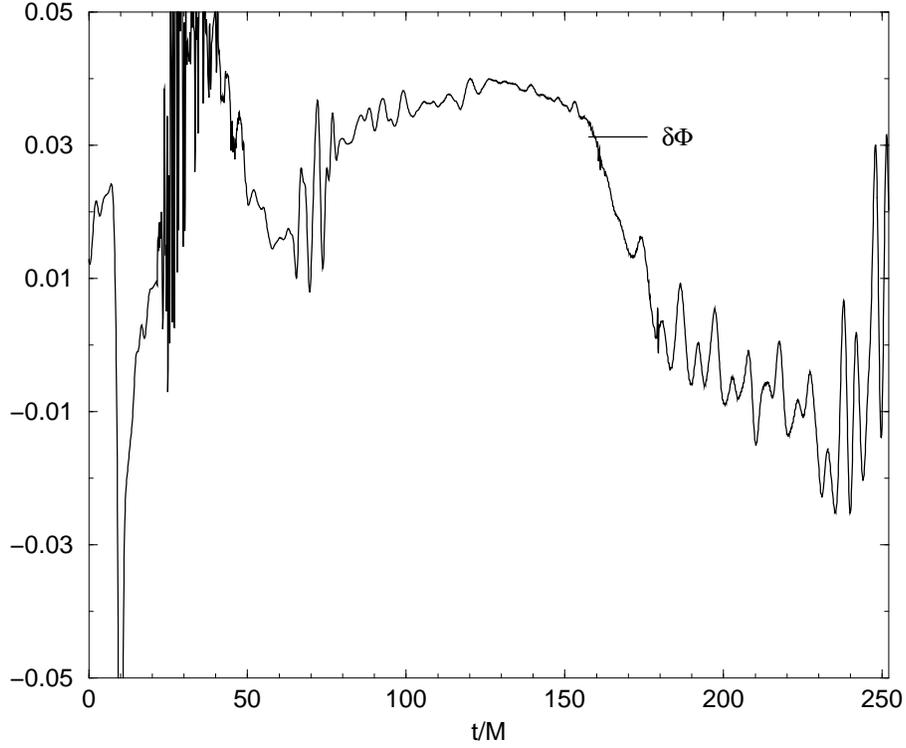}
    \caption{Plot of the time dependence of the difference in phase $\delta \Phi$,
 measured in radians, between the $(l=2,m=2)$ components of the characteristic
 waveform $\Psi(t)$ and the perturbative waveform $r\psi_4(t)/2$. The phase
 differences are less than 0.05 radians in the interval $40M < t < 250M$ beginning
 after the initial burst of junk radiation and extending into the late ringdown.
 The phase errors become large at late times when the numerical noise is
 comparable to the true signal amplitude and at very early times due the inability
 of the codes to accurately model the relatively high-frequency initial data
 burst. }
    \label{fig:phasediff}
 \end{figure}

Figure~\ref{fig:phasediff} plots the phase difference $\delta \Phi$ between the $(2,2)$
spherical harmonic components of $r \psi_4(t)/2$ and $\Psi(t)$, i.e
\begin{equation}
    \delta\Phi = \Phi[\Psi]-\Phi [ \psi_4],
\end{equation}
where e.g. $\Psi=  |\Psi|e^{i\Phi[\Psi]}$.
The phase difference is less than 0.05 radians in the interval $40M < t < 250M$  beginning
after the initial burst of junk radiation and extending into the late ringdown.
The phase errors become large at late times when the numerical noise in the
waveform is comparable to the true signal amplitude and at very early times due to
the inability of the codes to accurately model the relatively high-frequency
initial data burst.

\begin{figure}[htp] 
    \centering
    \includegraphics*[width=12cm]{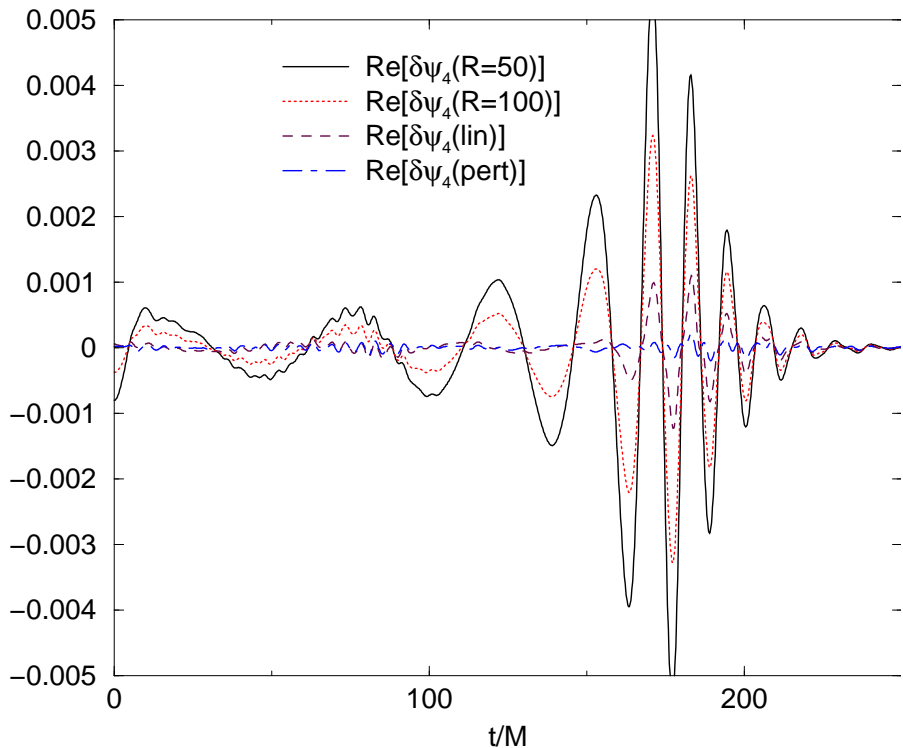}
    \caption{Plots of the time dependence of the difference $Re[\delta \psi_4]$
 between the $(l=2,m=2)$ components of the characteristic
 waveform $\Psi(t)$ and the Cauchy $\psi_4$ waveforms extracted at $R=50M$
 and $R=100M$, and their linear extrapolation to $R=\infty$ (denoted by ``lin'').
 For comparison, we also include the corresponding difference (denoted
 by ``pert'') using the perturbative extrapolation (\ref{eq:extrap}).
 The plots show the expected trend toward smaller
 errors as $R\to\infty$. Interestingly, perturbative extrapolation, which
 uses only the $R=100M$ extraction sphere, gives the smallest
 deviation.}
    \label{fig:lmagdiff}
 \end{figure}

\begin{figure}[htp] 
    \centering
    \includegraphics*[width=12cm]{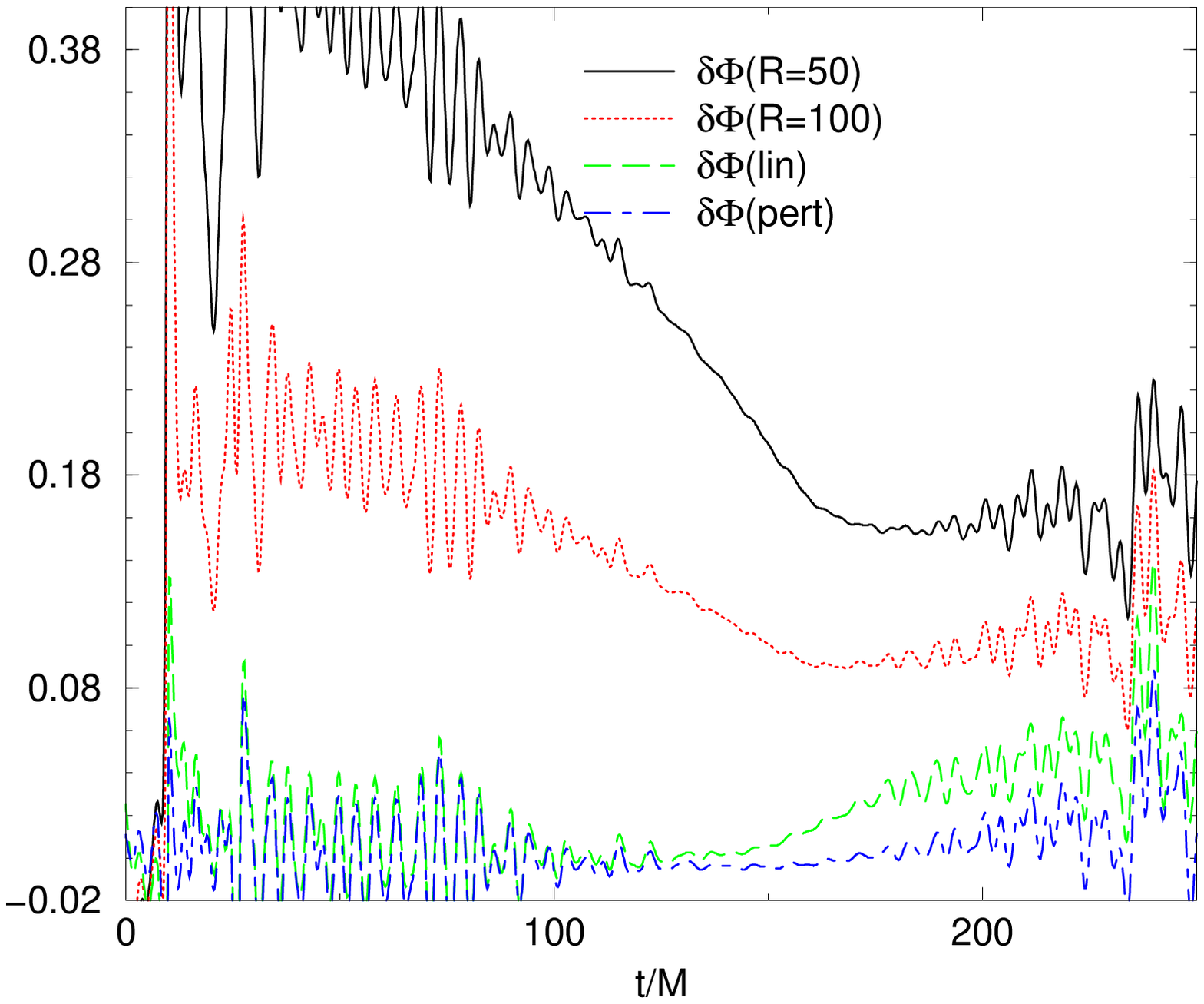}
    \caption{Plots of the time dependence of the phase difference $\delta\Phi $
 between the $(l=2,m=2)$ components of the characteristic
 waveform $\Psi(t)$ and the Cauchy $\psi_4$ waveforms extracted at $R=50M$
 and $R=100M$, and the corresponding linear extrapolation to
 $R=\infty$ (denoted by ``lin"). For comparison, we also include the
 corresponding $\delta \Phi$ (denoted by ``pert'') obtained by perturbative
 extrapolation (\ref{eq:extrap}). The plots show the expected trend toward smaller
 errors as $R\to\infty$. Interestingly, perturbative extrapolation, which
 uses only the $R=100M$ extraction sphere, gives the smallest deviation.}
    \label{fig:lphasediff}
 \end{figure}

Note that the magnitude and phase differences between $\psi_4$ extraction and characteristic extraction indicated in Figures~\ref{fig:magdiff} and \ref{fig:phasediff} are based upon the
perturbative extrapolation formula (\ref{eq:extrap}). It is also common practice
to compute $\psi_4$ at large radii and then
extrapolate the values to infinity, cf. the waveform comparisons in the report of
the Samurai project~\cite{samurai}. In carrying out the extrapolation,
the waveforms are translated by
$r^*$, where $r^* = r + 2 M \log (r/2M -1)$ is the tortoise coordinate obtained from the
areal radius of the extraction sphere $r$ and $M$ is the
ADM mass of the system~\cite{Boyle:2007ft}. 
Here we use a linear extrapolation based upon
waveforms at $R=50M$ and $R=100M$  to obtain an extrapolation $r\psi_4(lin)$
on $\scri^+$ that is accurate to order $O(1/R^2)$. The deviations from the characteristic
waveform are displayed in Figure~\ref{fig:lmagdiff}, where we plot $Re[\delta \psi_4(R=50M)]$,
$Re[\delta \psi_4(R=100M)]$ and $Re[\delta \psi_4(lin)]$ and in 
Figure~\ref{fig:lphasediff},  where we plot the phase differences $\delta\Phi(R=50M)$,
$\delta \Phi(R=100M)$ and $\delta\Phi(lin)$. The plots show how the deviations
decrease with extraction radius. Linear extrapolation considerably reduces the deviation
but it is interesting that perturbative extrapolation via (\ref{eq:extrap}), which is based
upon the single $R=100M$ result, gives the smallest deviation.

As we discuss next, such time domain comparisons can be of deceptive value for
gravitational wave data analysis. 

\section{Advanced LIGO accuracy standards}
\label{sec:crit}

It has been emphasized~\cite{linabuse} that the direct use of time domain errors
would be an abuse of the accuracy standards required of model waveforms to be
suitable for gravitational wave data analysis. The raw error envelopes $\delta
N(t)$, $\delta \Psi(t)$ and $\delta \psi_4(t)$  cannot be used to test whether the
accuracy standards are satisfied. Proper accuracy standards must take into account
the power spectral density of the detector noise $S_n(f)$, which is calibrated
with respect to the frequency domain strain $\hat h(f)$.  Consequently, the
primary accuracy standards must be formulated in the frequency domain in order to
take detector sensitivity into account. Fortunately, for the purpose of
calibrating waveforms from numerical simulations, it has been possible to
translate the frequency domain accuracy requirements into requirements on the time
domain $L_2$ error norms which meet all the needed criteria~\cite{lindbo,lindob}.

There are two distinct criteria for waveform accuracy. First, if the numerical
waveform were not sufficiently accurate then an unacceptable fraction of real
signals would pass undetected through the corresponding filter. Second, the
accuracy impacts on whether a detected waveform measures the physical properties
of the source, e.g. mass and spin, to a level commensurate with the accuracy of
the observational data. The accuracy standards for model waveforms have been
formulated to prevent these potential losses in the detection and scientific
measurement of gravitational waves.

For a numerical waveform with strain component $h(t)$, the time domain error is
measured by
\begin{equation}
     {\cal E}_0 = \frac {|| \delta h||}{|| h ||},
      \label{eq:strainerr}
\end{equation}
where $\delta h$ is the error in the numerical approximation and $||F||^2=\int dt
|F(t)|^2$, i.e. $||F||$ is the $L_2$ norm, which in principle should be integrated
over the complete time domain of the model waveform obtained by splicing a
perturbative chirp waveform to a numerical waveform for the inspiral and merger.

The error can also be measured in terms of time derivatives of the strain. In our
case, the first time derivative corresponds to the error in the news
\begin{equation}
        {\cal E}_1(Re N) = \frac {||\delta Re N ||}{|| Re N ||}  \, , \quad
        {\cal E}_1(Im N) = \frac {||\delta Im N ||}{|| Im N ||}
        \label{eq:newserr}
\end{equation}
and the second time derivative corresponds to the Weyl component error
\begin{equation}
        {\cal E}_2 (Re \Psi) = \frac {||\delta Re \Psi ||}{|| Re \Psi ||}  \, , \quad
        {\cal E}_2 (Im \Psi) = \frac {||\delta Im \Psi ||}{|| Im \Psi ||}  .
         \label{eq:weylerr}
\end{equation}
Here we measure $\delta N$ and $\delta \Psi$ according to (\ref{eq:Nerr}) and
(\ref{eq:Psierr}) and use the 3rd order Richardson extrapolations to compute $||Re
N||$, $||Im N||$, $||Re \Psi ||$and $||Im \Psi ||$. It is also of interest to
measure the ``error''
\begin{equation}
        {\cal E}_2(Re \delta \psi) = \frac {||Re \delta \psi_4||} {||Re \Psi ||} \, , \quad
         {\cal E}_2(Im \delta\psi) = \frac {||Im \delta \psi_4||} {||Im \Psi ||}
         \label{eq:delpsi}
\end{equation}
corresponding to the difference between perturbative and characteristic
extraction, where $\delta{\psi_4}$ is normalized according to (\ref{eq:psi4err}).

In~\cite{lindbo}, it was shown that sufficient conditions to satisfy data analysis
criteria for detection and measurement can be formulated in terms of any of the
error norms ${\cal E}_k=({\cal E}_0, {\cal E}_1,{\cal E}_2)$, i.e. in terms of the
strain, the news or the Weyl component.  The accuracy requirement derived
in~\cite{lindbo} for detection is
\begin{equation}
    {\cal E}_k \le C_k \sqrt{2\epsilon_{max}},
    \label{eq:Ndet}
\end{equation}
and the requirement for measurement is
\begin{equation}
    {\cal E}_k \le C_k \frac {\eta_c}{\rho}.
    \label{eq:Nmeas}
\end{equation}
Here $\rho$ is the optimal signal-to-noise ratio of the detector,
defined by
\begin{equation}
         \rho^2 = \int_0^\infty \frac{4|\hat h(f)|^2}{S_n(f)} df ;
\end{equation}
$C_k$ are dimensionless factors introduced in~\cite{lindbo}
to rescale the traditional signal-to-noise ratio $\rho$ in making the transition
from frequency domain standards to time domain standards;
$\epsilon_{max}$ determines the fraction of detections lost due to
template mismatch, cf. Eq. (14) of~\cite{lindob}; and
 $\eta_c\le 1$ corrects for error introduced in detector calibration.
These requirements  for detection and measurement, for either $k=0, 1,2$,
conservatively overstate the basic frequency domain requirements by
replacing $S_n(f)$ by its minimum value in transforming to the time domain.

The values of $C_k$ for the inspiral and merger of non-spinning, equal-mass black
holes have been calculated in~\cite{lindbo} for the advanced LIGO noise spectrum.
As the total mass of the binary varies from $0 \rightarrow \infty$, $C_0$ varies
between $.65 > C_0 >0$, $C_1$ varies between $.24 < C_1 < .8$ and  $C_2$ varies
between $0 < C_2 < 1$. Thus only the error ${\cal E}_1$  in the news  can satisfy
the criteria over the entire mass range. The error in the strain ${\cal E}_0$
provides the easiest way to satisfy the criteria in the low mass case
$M<<M_{\odot}$ and the  error in the Weyl component  ${\cal E}_2$ provides the
easiest way to satisfy the criteria in the high mass case $M>>M_{\odot}$.

We first concentrate on the error in the news, for which the accuracy requirement
for detection is
\begin{equation}
    {\cal E}_1 \le C_1 \sqrt{2\epsilon_{max}},
    \label{eq:Ndet2}
\end{equation}
and the requirement for measurement is
\begin{equation}
    {\cal E}_1 \le C_1 \frac {\eta_c}{\rho}.
    \label{eq:Nmeas2}
\end{equation}

Table~\ref{Errors}  gives values of several versions of the $ {\cal E}_1$ error
for the inspiral and merger of non-spinning, equal mass black holes described in
Sec.~\ref{sec:bbhwaveforms}. For practical purposes, the error norms were computed
over the time period of the simulation rather than for a complete model waveform
obtained by splicing to a post-Newtonian chirp waveform. Assuming that the
nonlinear error in the chirp waveform is small compared to the error in the
numerical waveform, the effect is to overestimate the error norms by
underestimating the denominators in (\ref{eq:strainerr}), (\ref{eq:newserr}) and
(\ref{eq:weylerr}). However, it has been pointed out in~\cite{hannam,pfeiffer}
that splicing to a chirp waveform can produce significant error unless the
numerical waveform extends to a large number of orbits, which can be
computationally prohibitive; otherwise, only for binary masses
$\gtrsim 100M_{\odot}$ is the splicing error negligible.

An advantage of the $ {\cal E}_1$ error norm based upon the
news function is that the  denominator in (\ref{eq:strainerr})
is directly related to the radiated energy. As a result, the factor by which
$ {\cal E}_1(N)$ is overestimated is ${\cal F}^{-1/2}$, where
\begin{equation}
       {\cal F}:= \frac   { \Delta E({\rm Numerical}) }
       {\Delta E({\rm Chirp})+\Delta E(\rm{Numerical})}
 \end{equation}
and $\Delta E$ denotes the energy radiated in the indicated time periods. The
total energy radiated during the post-Newtonian inspiral and merger can be
estimated from the difference between the final black hole mass $M_H$ and the mass
$M_0$ of the binary for a large initial orbit. The energy $ \Delta E({\rm
Numerical})$ radiated during the numerically modeled time period can be obtained
from the Bondi mass-loss formula. For the binary inspiral being considered here,
the final black hole mass is $M_H\approx 0.965187$; the initial mass of the system
at infinite separation (given by  the sum of the individual black hole masses) is
$M_0\approx 1.01447$; and   $ \Delta E({\rm Numerical})\approx 0.0346 $. This
leads to the fraction of energy $ {\cal F}\approx .702$ radiated during the
numerical period, or
\begin{equation}
     {\cal F}^{-1/2}\approx 1.19
     \label{eq:enfact}
\end{equation}
for the factor by which the $ {\cal E}_1$ errors in Table~\ref{Errors} are
overestimated. We re-emphasize that the values in the Table do not include
the error introduced by splicing the post-Newtonian and numerical waveforms.

Besides the values $ {\cal E}_1(N)$ of the numerical truncation error in the real
and imaginary part of the news function extracted at $R_E=20M$, $50M$ and $100M$,
Table~\ref{Errors}  includes the corresponding truncation error ${\cal
E}_1(N_\Psi)$ obtained from integrating $\Psi$ via (\ref{eq:npsi}). The Table also
includes the  modeling errors $ {\cal E}_1 (N_{\Delta R(20,100)})$ and $ {\cal
E}_1 (N_{\Delta R(50,100)})$ in the news which results from the differences
$N_{R=20}- N_{R=100}$ and $N_{R=50}- N_{R=100}$ obtained from extracting the
waveform at radii $R_E=50M$ and $R_E=20M$ compared with extraction at $R_E=100M$.
In computing these error norms, we integrate over the interval corresponding to
$t/M \ge 100$ in Fig.~\ref{fig:NewsR20R50R100} to eliminate effects of the initial
junk radiation.

\begin{table}[htdp]
\caption{Error norms of the $(2,2)$ spherical harmonic mode for the Bondi news
$N$, its counterpart $N_\Psi$  (obtained by time integral of the Weyl component
$\Psi$) and  for the differences $N_{\Delta R}$ comparing extraction at radii
$R_E=20M$ and $R_E=50M$ to extraction at $R_E=100M$. }
\begin{center}
\begin{tabular}{|c|c|c|}
   \hline
   $Variable$					   & 	$ Re $			& $Im $\\
   \hline 
   ${\cal E}_1(N)_{R=20}$	     		   &	$8.76\times10^{-4}$	&$8.74\times10^{-4}$\\
   ${\cal E}_1(N)_{R=50}$	     	 	   &	$2.62\times10^{-4}$	&$2.60\times10^{-4}$\\
   ${\cal E}_1(N)_{R=100}$	     	   &	$1.21\times10^{-4}$	&	$1.22\times10^{-4}$\\
   ${\cal E}_1 (N_\Psi)_{R=20}$	            & 	$1.08\times10^{-3}$	&	$1.12\times10^{-3}$\\ 
   ${\cal E}_1 (N_\Psi)_{R=50}$	            & 	$3.33\times10^{-4}$	&	$2.93\times10^{-4}$\\ 
   ${\cal E}_1 (N_\Psi)_{R=100}$          & 	$2.30\times10^{-4}$	&	$1.68\times10^{-4}$\\ 
   ${\cal E}_1 (N_{\Delta R(20,100)})$ &	$5.41\times10^{-3}$	&	$5.55\times10^{-3}$\\ 
   ${\cal E}_1 (N_{\Delta R(50,100)})$ &	$4.28\times10^{-3}$	&	$4.51\times10^{-3}$\\ 
      \hline
\end{tabular}
\end{center}
\label{Errors}
\end{table}%

Consider first the criterion for detection where we set $\epsilon_{max} =.005$,
which for advanced LIGO ensures less than a 10\% signal loss, a target which is
often adopted in LIGO searches for compact binaries~\cite{lindob}. For this
target, (\ref{eq:Ndet}) reduces to ${\cal E}_1  \le 0.1C_1$, or ${\cal E}_1  \le
.024$ for the low mass bound  $C_1\approx.24$. This criterion is easily satisfied
by all the error norms in Table~\ref{Errors}. Thus the advanced LIGO detection
criterion is satisfied by CCE waveforms obtained from either the news or Weyl
component throughout the entire binary mass range. In addition, the detection
criterion is unaffected by modeling errors introduced by choice of extraction
radius. Note that the ${\cal E}_1(N)$ and ${\cal E}_1 (N_\Psi)$ errors decrease
with larger extraction radius. This is expected since the truncation error
introduced by the characteristic evolution code depends upon the size of the
integration region between the extraction worldtube and $\scri^+$.

The criterion for measurement is more stringent. For a calibration  factor given
by the expected lower bound $\eta_{min} =0.4$ and  for the lower bound 
$C_1\approx.24$ corresponding to the small mass limit, (\ref{eq:Nmeas2}) reduces
to
\begin{equation}
     {\cal E}_1  \le C_1 \frac {\eta_c}{\rho}= \frac{ 9.6\times 10^{-2}}{\rho}.
    \label{eq:Nmeas3}
\end{equation}
For the most optimistic advanced LIGO signal-to-noise ratio, which is expected to
be $\rho \approx 100$ for the strongest and best tuned events, the requirement for
measurement is then $ {\cal E}_1 \le 9.6 \times 10^{-4}$.  Thus, comparing
(\ref{eq:Nmeas3}) to the values in Table~\ref{Errors},  the advanced LIGO
measurement criterion is satisfied throughout the entire binary mass range by the
numerical truncation error $ {\cal E}_1(N)$ in the CCE waveform obtained from the
news function. The ${\cal E}_1 (N_\Psi)$ error obtained from the Weyl component
for extraction radii $R_E\ge 50M$ also satisfy this full range of measurement
standards. The value of ${\cal E}_1 (N_\Psi)$ for
$R_E=20M$ would satisfy the full range of measurement standards
for $\rho< 100$ if reduced
by the factor ${\cal F}^{-1/2}$ given in (\ref{eq:enfact}). Note also that the
truncation error is being conservatively measured by the $O(\Delta^2)$ error
(\ref{eq:Psierr}), rather than the third order accurate error in the Richardson
extrapolated waveform.

These results can be compared with the measurement criterion for advanced LIGO
data analysis reported in the Samurai project~\cite{samurai}, which was also based upon
a non-spinning, equal-mass binary black hole inspiral and merger.
There it was found that the mismatch between perturbative waveforms obtained
using various Cauchy codes limited the measurement application  to
signal-to-noise ratios $\rho\lesssim 25$.  This is consistent with our experience,
and that reported in~\cite{reis2}, that the additional truncation error introduced by applying
CCE to a Cauchy simulation of a binary inspiral  is much smaller than the
numerical error resulting from the Cauchy code.

The values of  ${\cal E}_1 (N_{\Delta R})$ in Table~\ref{Errors} give an estimate
of the modeling error introduced by different choices of extraction radius. The
error ${\cal E}_1 (N_{\Delta R(50,100)})$, introduced by extraction at $R_E=50M$
as compared to $R_E=100M$, only satisfies the full range of measurement standards
for signal-to-noise ratios $\rho<21$, or  $\rho<25$ if (\ref{eq:enfact}) is taken
into account. This would cover the most likely advanced LIGO events. These
modeling errors primarily result from initialization effects which we have
discussed and which would be less significant in simulations with a higher number
of orbits. The results suggest that the choice of extraction radius should be
balanced between  a sufficiently large radius to reduce initialization effects and
a sufficiently small radius where the Cauchy grid is more highly refined and outer
boundary effects are better isolated. For the Cauchy grid setup in the present
case, there is a factor of 2 in refinement at $r= 50M$ compared to $r=100M$, which
for 8th order finite differencing has considerable impact on the error. Future
experiments with longer runs involving more orbits will supply valuable guidance
for optimizing the extraction radius.

Table~\ref{ErrorPsi4i} gives some pertinent ${\cal E}_2$ error norms for the
$(2,2)$ spherical harmonic component. Besides the numerical truncation error 
${\cal E}_2 (\Psi)$ obtained for  characteristic extraction at $R_E=20M$,
$R_E=50M$ and $R_E=100M$, the Table includes the error norm ${\cal
E}_2(\delta\psi)$ measuring the difference between perturbative and 
characteristic extraction, as defined in (\ref{eq:delpsi}), obtained at $R_E=50M$
and $R_E=100M$. The Table also includes the  modeling error $ {\cal E}_2
(\Psi_{\Delta R(50,100)})$ resulting from the difference $\Psi_{R=50}-
\Psi_{R=100}$ obtained using characteristic extraction at radii $R_E=50M$ and 
$R_E=100M$, as well as the corresponding error norm $ {\cal E}_2 (\psi_{4,\Delta
R(50,100)})$ resulting from the difference obtained using perturbative extraction
at $R_E=50M$ and  $R_E=100M$, i.e. 
\begin{eqnarray}
      {\cal E}_2 (Re \psi_{4,\Delta R(50,100)}) =
          \frac {||Re [(r\psi_4/2)|_{r=100M}- (r\psi_4/2)|_{r=50M}]||}
           {|| Re \Psi ||}  \\
     {\cal E}_2 (Im \psi_{4,\Delta R(50,100)}) =
          \frac {||Im [(r\psi_4/2)|_{r=100M}- (r\psi_4/2)|_{r=50M}]||} 
           {|| Im \Psi ||} .
\end{eqnarray}
All norms are again computed over the time interval $t/M \ge 100$ indicated in
Fig.~\ref{fig:NewsR20R50R100} to reduce effects of initial junk radiation.

Although the ${\cal E}_2$ norms are not effective for low mass binaries, they
give some useful information for comparing extraction at various radii and
comparing characteristic and perturbative extraction.  In the high mass limit for
which $C_2=1$ and with the same lower limits for $\epsilon_{max}$ and $\eta_c$ as
for the  ${\cal E}_1$ norms, the detection criterion (\ref{eq:Ndet}) reduces to
\begin{equation}
    {\cal E}_2 \le  \sqrt{2\epsilon_{max}}= 0.1
\end{equation}
and the measurement criterion (\ref{eq:Nmeas}) reduces to
\begin{equation}
    {\cal E}_2 \le  \frac {\eta_c}{\rho} =\frac {0.4}{\rho} .
    \label{eq:2meas}
\end{equation}

All the error norms in Table~\ref{ErrorPsi4i} satisfy the detection requirement
for this high mass limit.   The truncation errors ${\cal E}_2 (\Psi)$ decrease
with extraction radius as in the case of the ${\cal E}_1 (N_\Psi)$ errors. The
values at all three extraction radii satisfy the measurement requirement for the
most optimistic advanced LIGO signal-to-noise ratio $\rho=100$. These results are
consistent with the ${\cal E}_1 (N_{\Psi})$ error in Table \ref{Errors} obtained
by integrating $\Psi$.

The norms ${\cal E}_2(\delta \psi)$ measure the difference between characteristic
and perturbative extraction. The results in the Table show that this difference is
fairly independent of whether the waveforms are extracted at  $R_E=50M$ or
$R_E=100M$. In the high mass limit in which (\ref{eq:2meas}) is valid, these
errors impact the measurement criterion only for signal to noise ratios $\rho>59$
but they could be expected to be more significant for low mass binaries. Whether
the ${\cal E}_2(\delta \psi)$ error can be attributed to characteristic extraction
or to perturbative extraction cannot be decided from this single test and deserves
further investigation. A definitive answer would of course require knowledge of
the ``exact'' waveform.

The modeling error ${\cal E}_2(\psi_{4,\Delta R(50,100)})$, which results from
comparing perturbative extraction at $R_E=50M$ and $R_E=100M$, is considerably
larger than the corresponding modeling error ${\cal E}_2(\Psi_{\Delta R(50,100)})$
for characteristic extraction. This confirms the expectation that perturbative
extraction requires a large extraction radius. Both of these modeling errors are
substantial, which further emphasizes the importance of an optimal choice of
extraction radius.  

\begin{table}[htdp]
\caption{${\cal E}_2$ error norms for the $(2,2)$ spherical harmonic mode
 of the CCE waveform $\Psi$ obtained using extraction worldtube radii
 $R_E=20M$, $R_E=50M$ and $R_E=100M$ and the
 norms of the difference $\delta \psi$ between $\Psi$
 and the perturbative $\psi_4$ waveforms extracted at $50M$ and $100M$.
 We also tabulate the modeling error  ${\cal E}_2(\Psi_{\Delta R(50,100)})$
 resulting from the difference in extracting $\Psi$
 at $50M$ and $100M$, and the corresponding modeling error 
 ${\cal E}_2(\psi_{4,\Delta R(50,100)})$ for extraction via $\psi_4$
}
\begin{center}
\begin{tabular}{|c|c|c|}
   \hline
   $Variable$		 		& 	$ Re $				& $Im $\\
   \hline 
      ${\cal E}_2 (\Psi)_{R=20}$	&	$1.14\times10^{-3}$	&	$1.17\times10^{-3}$\\ 
      ${\cal E}_2 (\Psi)_{R=50}$	&	$4.04\times10^{-4}$	&	$3.53\times10^{-4}$\\
      ${\cal E}_2 (\Psi)_{R=100}$	&	$2.81\times10^{-4}$	&	$2.09\times10^{-4}$\\
      ${\cal E}_2 (\delta \psi)_{R=50}$	&	$5.09\times10^{-3}$	&	$5.08\times10^{-3}$\\ 
       ${\cal E}_2(\delta \psi)_{R=100}$	&	$6.81\times10^{-3}$	&$6.32\times10^{-3}$\\
       ${\cal E}_2(\Psi_{\Delta R(50,100)})	$&	$1.94\times10^{-2}$	&$1.91\times10^{-2}$\\
       ${\cal E}_2(\psi_{4,\Delta R(50,100)})	$&	$3.13\times10^{-2}$	&$3.14\times10^{-2}$\\ 
      \hline
\end{tabular}
\end{center}
\label{ErrorPsi4i}
\end{table}%
 
\section{Conclusion}

We have developed a new characteristic waveform extraction tool. Bugs and
inconsistencies in the previous version have been eliminated. The extracted
waveform from a binary black hole inspiral now shows clean convergence. We have
demonstrated that this allows the use of Richardson extrapolation to obtain third
order accurate waveforms whose numerical truncation error satisfies the advanced
LIGO standards for detection and measurement. Characteristic waveform extraction
from a binary black hole inspiral can now be obtained without any recourse to linearization
and from extraction radii as small as $R=20M$. The Cauchy interface has been
simplified in terms of a spectral decomposition. 

There are still elements where accuracy could be improved. Some of these, such as
more accurate start-up algorithms for the radial integrations at the extraction
worldtube and more accurate asymptotic limits at $\scri^+$, might be handled by
small modifications but others, such as extending the overall accuracy to 4th  or
higher order, would entail a more major overhaul of the underlying PITT code. This
is perhaps long overdue, but a proper treatment would require a better
understanding of the underlying mathematical problem. The well-posedness of the
gravitational worldtube-nullcone initial-boundary value problem upon which the
code is based has not yet been established. Only recently has well-posedness been
demonstrated for the corresponding nonlinear scalar wave problem~\cite{kwsc}. The
PITT code was developed in the early days of numerical relativity when
considerations of well-posedness did not arise in the formulation of Cauchy as
well as characteristic codes. The development of a stable characteristic code
involved  ``educated guesses''. Today, the numerical relativity community is more
aware of the benefits that a well-posed problem can bring.  Most important, a
proof of well-posedness of the continuum problem by means of energy estimates can
be converted to ensure stability of the corresponding finite difference problem by
the analogous estimates obtained by summation by parts. A new characteristic code
based upon this approach would be of great value. Of equal value would be the
implementation of Cauchy-characteristic matching (CCM), in which the
characteristic evolution is used to supply outer boundary data for the Cauchy
evolution. So far, CCM has only been successfully applied to a harmonic Cauchy
code in the linearized regime~\cite{harm}.

Although there is room for further improvement in the CCE tool presented here,
there also is pressing interest from several numerical relativity groups to apply
the tool to extract waveforms from binary black hole inspirals. The emerging
importance of this problem to the future of gravitational wave astronomy has
created an urgency to make characteristic waveform extraction widely available.
Simulations of binary black hole inspirals are too computationally expensive to be
carried out solely for the purpose of wave extraction tests. This would conflict
with the demands to apply computational resources to results of importance to 
gravitational wave astronomy and binary black hole astrophysics. However, the
extra computational expense of adding characteristic extraction to a Cauchy
simulation is fractionally small. For our tests, where we extracted twice as often
as required, the interpolation, decomposition, and saving of the metric data used
only $\approx6.9\%$ of the total simulation time. The application of
characteristic extraction to simulations of astrophysical importance will at the
same time provide a practical approach to improving the extraction tool by
comparing results obtained with different formulations, different  numerical
techniques and different grid specifications. In particular, our test results
emphasize the need for a better understanding of the optimal choice of extraction
radius, which would balance between the  discretization error in the Cauchy code,
the initialization error, the error originating at the outer Cauchy boundary and 
the relatively small discretization error from CCE.

We have demonstrated here how the module can be applied to the LazEv code, which
is a finite-difference BSSN code, to produce calibrated binary black hole
waveforms. We welcome applications to codes based upon other formulations of the
Einstein equations, e.g. the harmonic formulation, and based upon other numerical
methods, e.g. spectral methods. In particular, characteristic extraction offers a
way to unambiguously compare binary black hole waveforms obtained from the same
initial data using codes based upon different formulations of the Einstein
equations, different numerical techniques, different evolution gauges and
different methods of treating the internal singularities (by punctures or by
excision). Such comparisons would be of especial importance in the case of
precessing binaries composed of high spin black holes, where  the reliability of
perturbative extraction has not been extensively tested.

We have made the present characteristic waveform extraction tool publicly
available as part of the Einstein Toolkit~\cite{einstk}.


\begin{acknowledgments}
We thank L. Lindblom and C. Reisswig for many helpful discussions. B.~S.
acknowledges support from the Sherman Fairchild Foundation and NSF grants
PHY-061459 and PHY-0652995 to the California Institute of Technology. J.~W. 
acknowledges support from NSF grants PHY-0553597and PHY-0854623 to the University
of Pittsburgh.  Y.~Z. acknowledges support from NSF grants PHY-0722315,
PHY-0653303, PHY-0714388, PHY-0722703, DMS-0820923, PHY-0929114, PHY-0969855, and
NSF–CDI-1028087 and NASA grants 07-ATFP07-0158 and HST-AR-11763 to RIT and
computational resources provided by the Ranger cluster at TACC (Teragrid
allocations TG-PHY080040N and TG-PHY060027N) and by NewHorizons at RIT.  M.~C.~B. 
acknowledges support from NSF grant PHY-0969709 to the Marshall University and
computational resources provided by the Teragrid allocation TG-PHY090008. 
M.~C.~B. and B.~S. thank the University of Pittsburgh for hospitality during the
major part of this project. An essential component of this work is the PITT null
code, to which N.~T. Bishop, R. G{\' o}mez, R.~A. Isaacson, L. Lehner, P.
Papadopoulos and J. Welling have made major contributions.

\end{acknowledgments}

\appendix
\section{Code revision}
\label{sec:rev}

{\bf Revisions to the worldtube module:}

\begin{itemize}

\item  The numerical error in the previous version of the worldtube module did
not converge properly upon grid refinement. We have traced this problem to an
inconsistency in the startup algorithm for the integration of the characteristic
equations away from the extraction worldtube.  Data from the Cauchy code had
been used in an overdetermined manner to supply the integration constants for
the characteristic equations. As a result, the Cauchy evolution introduced
inconsistencies with the characteristic equations which degraded convergence of
the numerical error. We have revamped this start-up algorithm so that the
worldtube module now has clean second order accuracy with respect to grid size.

\item  We have found and corrected bugs which had been introduced
in the implementation of features designed to improve code performance.
In particular, in parallelizing the code using
the Cactus framework~\cite{cactus_web},  
a complex spin-weighted term in the evolution module was incorrectly declared
to be a real variable. In addition, it had been suggested that improved accuracy
could be obtained  by reducing second derivatives
in the angular directions to first order form by the use
of auxiliary variables~\cite{gomezfo}. In the process of doing so,  
values of certain variables in the subroutines for the data
at the extraction worldtube  were inadvertently interchanged
between the North and South stereographic
patches. The introduction of these bugs made the
resulting code inconsistent with the Einstein equations.
(From tracing through the code archive, we determined that
the bugs were introduced in 2002 or later so that they do not
affect the validity of results prior to 2002. C. Reisswig has informed us that
he recomputed some of the results in~\cite{reis1,reis2} using the corrected
code and found good qualitative agreement with the original results.)

\item The matching interface has been simplified by introducing a pseudospectral
decomposition of the Cauchy metric in the neighborhood of the extraction worldtube.
This provides more economical storage of the inner boundary data for the
characteristic code so that the waveform at $\scri^+$ can be obtained with small
computational burden compared to the Cauchy evolution.

\item  Interpolation error arises because the characteristic grid points do not lie
exactly on the extraction worldtube determined by the Cauchy coordinates. The
interpolation stencils change in a discontinuous way when the grid is refined.
Consequently, although this interpolation error  is second order in grid size, there
is a small stochastic component relative to the choice of grid. This can obscure the
results of convergence tests. We have reduced such sources of error so that
convergence tests can be used to validate the interface modules.

\item We have streamlined the start-up procedure at the extraction worldtube
by initializing the auxiliary variables (introduced to
remove second angular derivatives) directly in terms of the main variables. 

\item In previous applications of the extraction module, it was expedient to set
the characteristic data on the initial hypersurface to zero outside some radius.
This necessitated a transition region to obtain continuity with the initial Cauchy
data, which requires non-zero initial characteristic data at the extraction
worldtube. Here we initialize the data  by requiring that the Newman-Penrose 
component of the Weyl tensor intrinsic to the initial null hypersurface vanish,
i.e. $\psi_0 =0$. For a linear perturbation of the Schwarzschild metric, this
condition eliminates incoming radiation crossing the initial null hypersurface.
Since $\psi_0$ consists of a second radial derivative of the characteristic data,
this  condition allows both continuity at the extraction worldtube and the desired
asymptotic  falloff of the  characteristic data at infinity.

\end{itemize}

\medskip

{\bf Modifications of the PITT code:}

\begin{itemize}

\item  A source of error in characteristic evolution is the intergrid
interpolations arising from the stereographic patches used to coordinatize the
spherical cross-sections of the outgoing null hypersurfaces. The previous
version of the code used
two square stereographic patches centered about the North
and South poles, each overlapping the equator. This has now been modified by
shrinking the overlap region so that each patch has a circular boundary located
slightly past the equator, as is the practice in the use of stereographic grids
in meteorology~\cite{browning}. This eliminates the region near the corners of
the square patch where the numerical error was most troublesome. 
Angular numerical dissipation has also been introduced and shown to be
effective in controlling the short wavelength noise arising from the intergrid
interpolations across the stereographic patches. Tests show that
the resulting waveforms have smooth numerical error as
functions on the sphere~\cite{strat}.

Characteristic codes based upon a  six patch covering of the
sphere~\cite{reisswig,roberto} offer
the potential for better accuracy but they have not yet been developed to handle
waveform extraction. See~\cite{strat} for a comparison of the six patch and the
stereographic approaches on a test problem.

\item The accuracy of the angular derivatives has been increased to a 4th order
finite difference approximation, as opposed to the 2nd order accuracy in the
original code. The radial derivatives and time integration remain
second order accurate.

\item Some of the differential equations governing propagation along the
characteristics become degenerate at $\scri^+$ and affect the accuracy
of asymptotic quantities such as the Bondi news function. The correct asymptotic
behavior has now been incorporated into the finite difference approximation in order to
increase accuracy. In addition,
the accuracy of certain one-sided finite difference approximations
necessary at $\scri^+$ has also been improved.

\item In addition, the code has been extended to supply the waveform at $\scri^+$ in
terms of the radiative component of the Weyl tensor as well as the Bondi news
function. For tests in the linearized regime, extraction via the Weyl
tensor was found to be slightly more accurate than via the news
function when large gauge effects are introduced in the characteristic
coordinates~\cite{strat}. On the other hand, the higher derivatives involved in
computing the Weyl tensor lead to less smoothness in the numerical error.
Overall, the two methods are competitive.

\end{itemize}

\end {document}